\def\equationautorefname~#1\null{Equation (#1)\null}
\def\sectionautorefname~#1\null{Section #1\null}
\def\subsectionautorefname~#1\null{Section #1\null}
\def\subsubsectionautorefname~#1\null{Section #1\null}
\def\figureautorefname~#1\null{Figure #1\null}

\documentclass[twocolumn]{aastex63}
\usepackage{lineno}
\usepackage{amsmath}
\usepackage{xparse}
\usepackage{hyperref}
\usepackage{longtable}
\usepackage[flushleft]{threeparttable}
\usepackage[titletoc,title]{appendix}
\usepackage{graphicx}% Include figure files
\usepackage{bm}% bold math

\newcommand*{\ar}{\autoref}
\newcommand*{\ct}{\citet}
\newcommand*{\ctp}{\citep}
\newcommand*{\cta}{\citealp}
\newcommand*{\tcb}{\textcolor{black}}

\received{}
\revised{}
\accepted{\today}
\graphicspath{{./}{figures/}}
\begin{document}
%%%========================================================
\title{Nebular spectra of kilonovae with detailed recombination rates - I. Light r-process composition} 
\correspondingauthor{Smaranika Banerjee}
\email{smaranika.banerjee@astro.su.se}

\author[0000-0002-0786-7307]{Smaranika Banerjee}
\affiliation{The Oskar Klein Centre, Department of Astronomy, Stockholm University, AlbaNova, SE-10691 Stockholm, Sweden}

\author[0000-0001-8005-4030]{Anders Jerkstrand}
\affiliation{The Oskar Klein Centre, Department of Astronomy, Stockholm University, AlbaNova, SE-10691 Stockholm, Sweden}

\author{Nigel Badnell\textsuperscript{\dag}}
\affiliation{Department of Physics, University of Strathclyde, Glasgow G4 0NG, United Kingdom.}

\thanks{\dag\ Deceased}
% \author*{Nigel Badnell}
% \author{Nigel Badnell\footnote[2]{Deceased}}
% \author{John Smith\thanks{Deceased}}

\author[0000-0001-5255-0782]{Quentin Pognan}
\affiliation{Max-Planck-Institute for Gravitational Physics (Albert Einstein Institute), Am M\"{u}hlenberg 1, Potsdam-Golm, 14476, Germany}

\author[0009-0005-3148-513X]{Niamh Ferguson}
\affiliation{Department of Physics, University of Strathclyde, Glasgow G4 0NG, United Kingdom.}

\author[0000-0002-6224-3492]{Jon Grumer}
\affiliation{Theoretical Astrophysics, Department of Physics and Astronomy, Uppsala University, Box 516, SE-751 20 Uppsala, Sweden}

%========================================================
%%abstract
%========================================================
\begin{abstract}
To investigate spectra of kilonovae in the NLTE phase ($t\gtrsim$ 1 week),
we perform atomic calculations for dielectronic recombination (DR) rates for the light $r$-process elements Se ($Z = 34$), Rb ($Z = 37$), Sr ($Z = 38$), Y ($Z = 39$), and Zr ($Z = 40$)
using the \texttt{HULLAC} code. For the different elements, our results for the DR rate coefficients for recombining from the ionization states of II to I, III to II, and IV to III vary between
$2\times10^{-12} - 5\times10^{-11} \, \rm cm^3\,s^{-1}$, $10^{-13} - 5\times10^{-11} \, \rm cm^3\,s^{-1}$
and $2\times10^{-15} - 10^{-11} \, \rm cm^3\,s^{-1}$, respectively, at a temperature of $T = 10,000$ K.
Using this new atomic data (DR), we study the impact on kilonova model spectra at phases of  $t = 10$ days and $t = 25$ days {after the merger} using the spectral synthesis code \texttt{SUMO}.
Compared to models using the previous treatment of recombination as a constant rate, the new models show significant changes in ionization and temperature, and correspondingly, in emergent spectra. 
With the new rates, we find that Zr ($Z = 40$) plays a yet more dominant role in kilonova spectra for light $r$-process compositions.
Further, we show that previously predicted mid-infrared (e.g. [Se III] 4.55 $\mu$m) and optical (e.g. Rb I 7802, 7949 Å) lines {weaken} in the new model.
Instead [Se I] $5.03\ \mu$m emerges as a signature. 
These results demonstrate the importance of considering the detailed microphysics for modelling and interpreting the late-time kilonova spectra. 
%which is the ideal target for the \textit{James Webb Space Telescope}.
\end{abstract}

%==============================================================
%==============================================================

\section{Introduction} \label{sec:intro}

It has long been hypothesized that $r$-process elements are synthesized in neutron star mergers
(e.g., \cta{Lattimer74, Eichler89, Freiburghaus99, Korobkin12, Wanajo14}).
The radioactive decays of the freshly synthesized heavy elements give rise to a transient at
ultraviolet, optical, and near-infrared wavelengths called a kilonova (e.g., \citealt{Li98, Kulkarni05, Metzger10}).
The first kilonova, AT2017gfo (e.g., \cta{Coulter17, Yang17,Valenti17, Cowperthwaite17, Smartt17, Drout17, Utsumi17})
was observed following the gravitational wave event GW170817 \citep{Abbott17a}, opening up a new avenue of multi-messenger astrophysics.

Several efforts have been made to interpret the light curves of the kilonova observed (AT2017gfo) starting from as early as an hour
(e.g., \cta{kasen17, Tanaka17, Shibata17, Perego17, Rosswog18, Kawaguchi18, Wollaeger19, Smaranikab20, Wollaeger21, Korobkin21, Smaranikab22, Smaranikab24}),
confirming the prediction about the synthesis of $r$-process elements in neutron star mergers.
Furthermore, the modelling of AT2017gfo spectra has indicated signatures of several heavy elements; 
a few of the examples include Sr ($Z = 38$, \cta{Watson19}), {Y ($Z = 39$, \citealt{Sneppen24})}, and Ce ($Z = 58$, \cta{Domoto22}). 

The works mentioned are based on the modelling and interpretation of the early photospheric-phase spectra.
At this phase, the bulk of the ejecta are optically thick, and the kilonova is in the diffusion phase.
In this phase, an assumption of local thermal equilibrium (LTE) is used to model the kilonova light curves and spectra.
However, after about a week or so, the diffusion phase ends and the kilonova enters its tail phase. 
The assumption of equilibrium breaks down and photospheric scattering lines now start to give way to emission lines,
even though there is still significant line opacity and associated fluorescence \citep{Pognan23}. 
The work on this phase requires considering non-LTE (NLTE) and has begun more recently (e.g., \cta{Hotokezaka21, Hotokezaka22, Pognan22a, Pognan23, Hotokezaka23}).
The NLTE spectral modelling of kilonovae has already indicated a few elemental signatures, such as Se ($Z = 34$, \cta{Hotokezaka22}), Te ($Z = 34$, \cta{Hotokezaka23}), and Rb ($Z = 37$, \cta{Pognan23}). 

To model kilonovae in the NLTE phase, the ionization and excitation populations need to be determined by solving rate equations (e.g., \cta{Jerkstrand2017,Hotokezaka22}). 
Such calculations require the rates and the cross-sections of different processes for the $r$-process elements. Such data include:
(1) Energy levels and spontaneous radiative transition rates (A-values),
(2) thermal collisional bound-bound rates,
(3) non-thermal collisional bound-free cross-sections,
(4) photo-ionization (PI) cross-sections, and
(5) (thermal) recombination rates.

The availability of experimental and theoretical calculations
for such atomic data is limited for the $r$-process elements, and previous works to model KN spectra have often involved new calculations of the needed atomic data.
For example, \ct{Hotokezaka21} calculate the energy levels and A-values
by using \texttt{GRASP2K} \ctp{Grasp19} for Nd (Z=60) at ionization states II - IV.
In \ct{Pognan23}, energy levels and A-values are calculated with the
\texttt{Flexible Atomic Code} (\texttt{FAC}, \cta{Gu08}) for the elements Ga ($Z = 31$) - U ($Z = 92$) for ionization states I - IV.

In the nebular phase, the ionization occurs mainly by non-thermal collisions by the high-energy electrons created in the downscattering cascade of radioactive decay particles, and by photoionization (PI).
Recombination proceeds via direct radiative (RR) and resonant dielectronic recombination (DR) processes.
Hence, DR and RR rates, as well as PI cross-sections, are crucial for accurate modelling of the ionization state in the NLTE phase.
Such recombination rates and PI cross-sections for $r$-process elements, especially at the low ionization states applicable for kilonova, are scarce, with only a few existing works.
For example, \ct{Preval19} calculate the recombination rate coefficients of the lower ionized W ($Z = 74$, ionization states II to XIV). Furthermore, \cite{Sterling11} and \cite{Sterling11b} calculate the recombination and PI cross-sections of low charged Se ($Z = 34$) and Kr ($Z = 36$) ions (up to ionization states of VI).  However, such sparse works are not enough to calculate accurate spectra of kilonovae at nebular phases as the BNS merger ejecta contains mixture of heavy elements.
%no systematic work covering the majority of $r$-process elements exists.

In the work of \ct{Pognan23}, a constant (temperature-independent) total recombination rate of $10^{-11}\, \rm cm^3 \, s^{-1}$ is used, and a hydrogenic treatment for PI cross-sections.
%Additionally, a simplistic hydrogenic treatment for PI cross-sections is used.
Conversely, \citet{Hotokezaka21} perform atomic calculations to determine DR rates only for Nd ions (stages II - IV) and follow the analytical treatment of \ct{Axelrod80} for RR rates.

%, for several $r$-process elements.}
%including both DR and RR contributions, for several $r$-process elements.
%determine the total recombination rate coefficients,
%We also compute the PI cross-sections as the inverse process of RR (through the Milne relations).
%Since calculation of atomic rates for all $r$-process elements is a very large endeavor, we focus initially on a subset of light $r$-process elements:
In this work, {we aim to improve the situation for the lack of the atomic data for the calculation of the nebular spectra of kilonovae. As this is a very large endeavor, we first perform atomic calculations to investigate the DR contributions} for a subset of light $r$-process elements \tcb{(i.e., the elements with $Z < 56$)}:
Se ($Z = 34$), Rb ($Z = 37$), Sr ($Z = 38$), Y ($Z = 39$), and Zr ($Z = 40$).
These elements are chosen because they have demonstrated potential to produce strong spectral signatures in kilonovae \ctp{Hotokezaka22,Gillanders22, Pognan23}. We consider recombination to I - III ions (neutral to doubly ionized), as these are most relevant during the nebular phase (e.g., \cta{Pognan23}).
Using this new data, we then study how kilonova NLTE models are impacted using the \texttt{SUMO} spectral synthesis code.

The paper is structured as follows.
We discuss the details of the atomic data requirements and our method of calculation in \ar{sec:at_data}.
We provide the results for our atomic calculations in \ar{sec:rr_dr}.
We discuss our spectral synthesis results in \ar{sec:radtr}. 
We finally summarize our conclusions in \ar{sec:conclusion}.

%========================================================
\begin{table*}[]
\centering
\caption{Configurations used to perform atomic calculations. The bold configurations are used as ground configurations and optimization.}
\label{tab:config}
\resizebox{\textwidth}{!}{%
\begin{tabular}{ccl}
\hline
Element & Ion & \multicolumn{1}{c}{Configurations}                      \\ \hline
Se      & I   & \begin{tabular}[c]{@{}l@{}}$\bf{4s^2 4p^4}$, $\bf{4s^2 4p^3 4d}$, $\bf{4s^2 4p^3 4f}$, $4s^2 4p^3 nl$ ($n\le 10$ for $l\le3$, $n\le 8$ for $l\le5$), \\ $4s^2 4p^2 4d 5p$,  $4s 4p^5$, $4s 4p^4 5p$, $4s 4p^3 5p^2$, $4p^6$\end{tabular}                                \\ 
Se      & II  & \begin{tabular}[c]{@{}l@{}}$\bf{4s^2 4p^3}$, $\bf{4s 4p^4}$, $4s^2 4p^2 nl$ (for $l\le2$, $n\le9$, $2< l \le 5$, $n\le7$)\end{tabular}                                                                                       \\
Se      & III & \begin{tabular}[c]{@{}l@{}}$\bf{4s^2 4p^2}$, $\bf{4s 4p^3}$, $4s^2 4p  nl$ ($n \le 13$ for $l = 0$, $n\le 11$ for $0< l \le 2$, $n\le 9$ for $2< l \le 4$,  \\ $n\le 10$ for $l = 5$)\end{tabular}                            \\
Se      & IV  & \begin{tabular}[c]{@{}l@{}}$\bf{4s^2 4p}$, $4s^2 nl$ ($n\le 9$ for $0 < l \le 2$, $n\le 7$ for $2< l \le 5$), $4p^3, \, 4s \, 4p^2$\end{tabular}                                                                                    \\ \hline
Rb      & I   & \begin{tabular}[c]{@{}l@{}}$\bf{4p^6 5s}$, $\bf{4p^6 5p}$, $\bf{4p^6 4d}$, $\bf{4p^6 4f}$, $4p^6 nl$ ($n \le 10$ for $l\le 2$, $n \le 8$ for $l\le5$),\\ $4p^5 5s^2$, $4p^5 5s nl$ ($n \le 7$ for $l\le 2$, $n \le 5$ for $l= 3$) \end{tabular}                                                        \\
Rb      & II  & \begin{tabular}[c]{@{}l@{}}$\bf{4p^6}$, $\bf{4p^5 5s}$, $\bf{4p^5 5p}$, $\bf{4p^5 4d}$, $\bf{4p^5 4f}$, $4p^5 nl$ ($n \le 10$ for $l\le 2$, $n \le 8$ for $l\le5$)       \end{tabular}                                                                                                                                                                                       \\
Rb      & III & \begin{tabular}[c]{@{}l@{}}$\bf{4p^5}$, $\bf{4p^4 4d}$, $\bf{4p^4 5p}$, $\bf{4p^4 5s}$, $\bf{4p^4 4f}$, $4p^4 nl$ ($n \le 10$ for $l\le 2$, $n \le 8$ for $l\le5$)      \end{tabular}                                                                                                                                                                                     \\
Rb      & IV  & \begin{tabular}[c]{@{}l@{}}$\bf{4p^4}$, $4p^3 nl$ ($n \le 10$ for $l\le 2$, $n \le 8$ for $l\le5$)     \end{tabular}    \\ \hline
%
%-------
Sr      & I   & \begin{tabular}[c]{@{}l@{}}$\bf{5s^2}$, $\bf{5s 5p}$, $\bf{5s 4d}$, $5s nl$ ($n \le 10$ for $l\le 2$, $n \le 8$ for $l\le 5$),
$4d nl$ ($n \le 10$ for $l\le 1$, $n \le 8$ for $l\le 5$), \\ $5p nl$ ($n \le 10$ for $l\le 1$, $n \le 8$ for $l\le 4$, $n \le 7$ for $l=5$), $5s^2$ \end{tabular}   \\
Sr      & II  & \begin{tabular}[c]{@{}l@{}}$\bf{4p^6 5s}$, $\bf{4p^6 4d}$, $\bf{4p^6 5p}$, $\bf{4p^6 4f}$, $4p^6 nl$ ($n \le 8$ for $l\le 3$, $n \le 6$ for $l\le 5$),
\\ $4p^5 5s nl$ ($n \le 8$ for $l\le 2$, $n \le 6$ for $l\le 5$), $4p^5 5s^2$, $4p^5 4d^2$, $4p^5 5p^2$, $4p^5 4f^2$\end{tabular}   \\
Sr      & III & \begin{tabular}[c]{@{}l@{}}$\bf{4s^2 4p^6}$, $4p^5 4s^2 nl$ ($n \le 10$ for $l\le 2$, $n \le 8$ for $l\le 5$), \\$4p^4 4s^2 nl^2$ ($n \le 10$ for $l\le 2$, $n \le 8$ for $l\le 5$)\end{tabular}  \\
Sr      & IV  & \begin{tabular}[c]{@{}l@{}}$\bf{5s^2 4p^5}$, $4s 4p^6$, $4s^2 4p^4 nl$ ($n \le 10$ for $l\le 2$, $n \le 8$ for $l\le 5$),\\ $4s^1 4p^5 nl$ ($n \le 10$ for $l< 2$, $n = 13$ for $l = 2$, $n \le 8$ for $l\le 5$)\end{tabular}  \\ \hline
%
%----------
Y       & I   & \begin{tabular}[c]{@{}l@{}}$\bf{4d 5s^2}$, $\bf{4d^2 5s}$, $\bf{4d 5s 5p}$, $\bf{5s^2 5p}$, $5s^2 6p$, $4d 5s nl$ ($n \le 10$ for $l\le 2$, $n \le 8$ for $l\le 5$),\\ $4d^2 nl$ ($n \le 10$ for $l\le 2$, $n \le 8$ for $l\le 5$),
$4d^3$, $5s^2 5d$, $5s^2 7s$, $5p^2 4d$\end{tabular} \\
Y       & II  & \begin{tabular}[c]{@{}l@{}}$\bf{5s^2}$, $\bf{4d^2}$, $\bf{4d 5s}$, $4d 5p$, $5s nl$ ($n \le 10$ for $l\le 2$, $n \le 8$ for $l\le 5$)\\ $4d nl$ ($n \le 10$ for $l< 2$, $n \le 8$ for $l\le 5$)\end{tabular}   \\
Y       & III & \begin{tabular}[c]{@{}l@{}}$\bf{4p^6 4d}$, $\bf{4p^6 5s}$, $\bf{4p^6 5p}$, $\bf{4p^6 4f}$, $4p^6 nl$ ($n \le 8$ for $l\le 3$, $n \le 6$ for $l\le 5$),
\\ $4p^5 5s nl$ ($n \le 8$ for $l\le 3$, $n \le 6$ for $l\le 5$), $4p^5 5s^2$, $4p^5 4d^2$, $4p^5 5p^2$, $4p^5 4f^2$\end{tabular}   \\
Y       & IV  & \begin{tabular}[c]{@{}l@{}}$\bf{4p^6}$, $4p^5 nl$ ($n \le 10$ for $l\le 2$, $n \le 8$ for $l\le 5$)    \end{tabular} \\ \hline
%
%-----------
Zr      & I   & \begin{tabular}[c]{@{}l@{}}$\bf{4d^2 5s^2}$, $\bf{4d^3 5s}$, $\bf{4d^2 5s 5p}$, $4d^2 5p^2$, $4d 5s^2 5p$, $4d 5s 5p^2$, $4d 5p^3$, $4d^4$,
\\ $4d^3 nl$ ($n \le 10$ for $l\le 2$),$4d^2 5s nl$ ($n \le 10$ for $l= 0$, $n \le 9$ for $l\le 5$)\end{tabular}      \\
Zr      & II  & \begin{tabular}[c]{@{}l@{}}$\bf{4d^2 5s}$, $\bf{4d 5s^2}$, $\bf{4d^2 5p}$, $\bf{4d 5p^2}$, $\bf{4d 5s 5p}$, $\bf{4d^3}$,
\\ $4d 5s nl$ ($n \le 10$ for $l\le 2$, $n \le 8$ for $l\le 5$)\end{tabular}   \\

Zr      & III & \begin{tabular}[c]{@{}l@{}}$\bf{4d^2}$, $5p^2$, $4d \,nl$ ($n\le 10$ for $l\le 2$, $n \le 8$ for $l \le 5$),
\\ $5s \,nl$ ($n\le 8$ for $l\le 2$, $n \le 6$ for $l \le 5$)\end{tabular}   \\
Zr      & IV  & \begin{tabular}[c]{@{}l@{}}$\bf{4d}$, $nl$ ($n \le 8$ for $l \le 5$)   \end{tabular}  \\ \hline
\end{tabular}
}
\end{table*}

%========================================================
\section{Atomic data for nebular spectra} \label{sec:at_data}
%========================================================
% In addition, we use the analytical estimate of RR rates for Fe calculated in \ct{Axelrod80}.
%For the PI cross-section, 
%========================================================
\subsection{HULLAC} \label{subsec:hu}
%========================================================

For the atomic {structure} calculations {for DR rates}, we use \texttt{HULLAC} (Hebrew University Lawrence Livermore Atomic Code, \cta{Bar-shalom01}).
\texttt{HULLAC} is a set of programs that uses the same set of wavefunctions to calculate the different processes with the same level of accuracy.
The performance of \texttt{HULLAC} is focused on the ionized heavy elements with several open shells.
This is one of the major reasons for using this code since the main purpose of this work is to perform atomic calculations for such ions.
The theoretical framework is outlined in previous works \ctp{Bar-shalom01,Tanaka18, Smaranikab22, Smaranikab24}.
Hence, we only briefly summarize some of the key theoretical aspects of this code here.

\texttt{HULLAC} calculates a set of fully relativistic orbitals to determine the energy levels and radiative transition probabilities.
The main theoretical framework of \texttt{HULLAC} revolves around using the first order perturbation theory with a central field potential,
which includes both nuclear field and the spherically averaged electron-electron interaction.
The zero order orbitals are the solution of the single electron Dirac equation. \texttt{HULLAC} uses parametrized central field potential.
The free parameter of the distribution is determined by the minimization of the first-order configuration averaged energies for the selected configurations.
For the continuum orbitals, the phase amplitude equation is solved for the parametric potential for the bound electrons with an asymptotic correction.

We use the \textit{level mode} of \texttt{HULLAC} to perform the calculations of the {DR rate coefficients}. %recombination cross sections.
The configurations used for the \texttt{HULLAC} calculations are listed in \ar{tab:config}. 
Note that we do not focus extensively on optimizing the atomic data, as our primary goal is to investigate how
the detailed recombination rate coefficients and photo-ionization cross-sections will modify the kilonova nebular spectra,
when used, in comparison to the constant rates used before \ctp{Pognan23}.
Further optimization and fine tuning of the data is within the scope of the future work.

%========================================================
%\subsection{Method} \label{subsec:at_meth}
%=======================================================

%Below we discuss in more detail the calculations for rates of direct RR, resonant DR, and PI cross-sections. 

%=======================================================
\subsection{Dielectronic recombination} 
%=======================================================
{One of the routes} via which recombination occurs is DR, which is a resonant process.
In this case, the kinetic energy of the free electron excites a bound electron to move to a bound excited state,
and in the process, produces a short lived doubly excited autoionizing state.
The doubly excited system either autoionizes back to the original configuration, 
or decays to form a (recombined) singly excited bound state.
If the second process occurs, the DR process has {happened}.
Since the energy of the electron must be the same as the energy difference between the two states $X_{p}^{(i+1)+}$ and $X_a^{i+}$, 
this is a resonant process. This two-step resonant DR process can be written as:
\begin{equation}
X_p^{(i+1)+} + e^- \rightarrow X_a^{i+} \rightarrow X_b^{i+} + h\nu,
\end{equation}
where $a$ represents the autoionizing state.

DR rate coefficients can be determined as the dielectronic capture rate times the fraction that decays into bound states (branching ratio).
Assuming the DR is dominated by the population in the ground level of the recombining ion ($X_p^{(i+1)+}$),
the DR rate through autoionization state $a$ is calculated as \ctp{Burgess64, Nussbaumer83}:
\begin{equation}
\alpha_{\rm DR} (p,a, T_e) = \dfrac{N_{\rm S}(X_a ^{i+})}{N_e N_{\rm S}(X_a ^{(i+1)+})}\dfrac{\sum_{j}A_{aj}\sum_{c}\Gamma_{ac}}{\sum_{c}\Gamma_{ac} +\sum_{k}A_{ak}},
\label{eq:burgess}
\end{equation}
where $A_{aL}$ is the radiative transition rate from state $a$ to the bound state $L$, $\Gamma_{ac}$ is the autoionization rate between the autoionizing state $a$ and the continuum state $c$. The $j$-summation in \ar{eq:burgess} is over all the levels that are stable against autoionization, whereas the $k$-summation is over all the lower levels.
The $c$-summation is over all the energetically accessible continuum levels.
The total DR rate coefficient is calculated by the summation over the individual autoionizing levels. 
Finally, $N_S$ is the number density at thermo-dynamic equilibrium, with $N_e$ being the electron number density. 
The level population term is given as \ctp{Burgess64, Nussbaumer83}
\begin{equation}
\dfrac{N_{\rm S}(X_a ^{i+})}{N_e N_{\rm S}(X_a ^{(i+1)+})} = \dfrac{g_a}{2g_p}\bigg(\dfrac{h^2}{2\pi m_ekT_e}\bigg)^{3/2} e^{-E_{\rm th}/kT_e}. 
\end{equation}
Here $g_{a}$ and $g_{p}$ are the statistical weights of the autoionizing state $a$ and the ground state $p$ of the recombining ion, $E_{\rm th}$
is the threshold energy between the autoionization state and the higher ion state, and the other symbols have their usual meaning.

At nebular temperatures, in the low density plasma of the kilonova ejecta, mainly the ground term and the first few excited states of the ions are significantly populated. 
Hence, the autoionizing states only slightly above the ionization threshold can significantly contribute to recombination.
Therefore, we choose to resolve the energy levels within 2 eV of the ionization threshold for our calculation.

%========================================================
\begin{figure*}[t]
  \begin{tabular}{c}
 
   \begin{minipage}{0.5\hsize}
      \begin{center}
       
        \includegraphics[width=\linewidth]{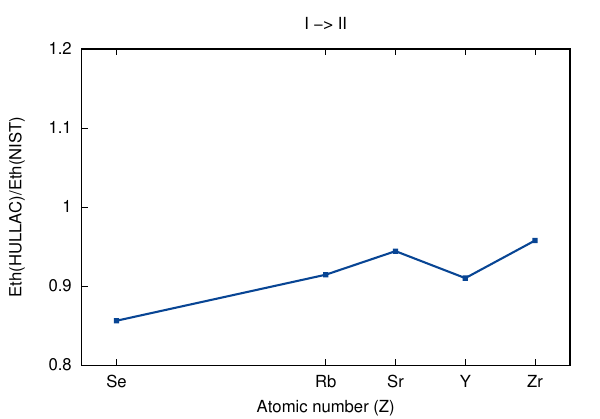}
      \end{center}
    \end{minipage}
    
    \begin{minipage}{0.5\hsize}
      \begin{center}
        
        \includegraphics[width=\linewidth]{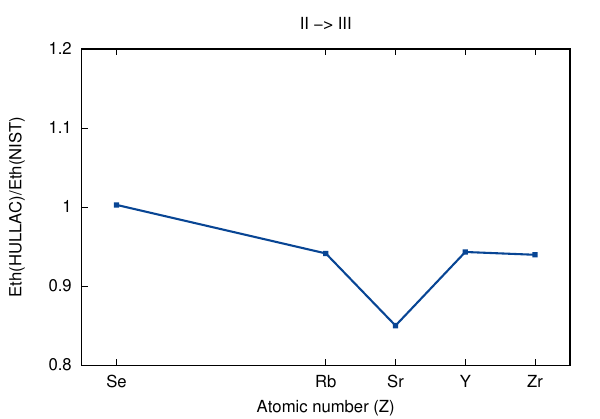}
      \end{center}
    \end{minipage}
    \\
    \begin{minipage}{0.5\hsize}
      \begin{center}
        
        \includegraphics[width=\linewidth]{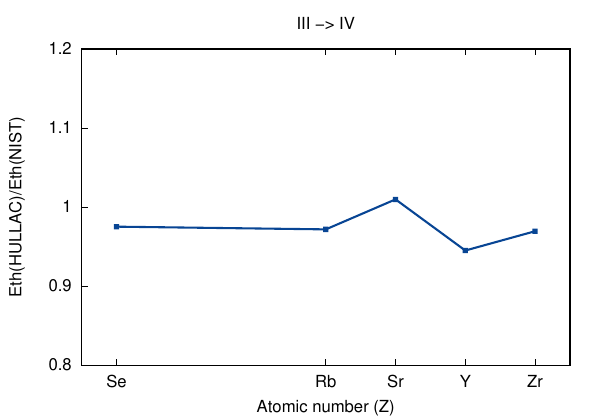}
      \end{center}
    \end{minipage}

\end{tabular}
  \caption{The ratio between the (ground state) ionization threshold energies obtained from \texttt{HULLAC} to those provided in the NIST database \ctp{NIST20}.
  The results show good agreement across all the different elements considered.}
  \label{fig:eth}
\end{figure*}

%========================================================
% figure rr 
%========================================================
% \begin{figure*}[t]
%   \begin{tabular}{c}
 
%    \begin{minipage}{0.5\hsize}
%       \begin{center}
%        %\includegraphics[width=\linewidth]{./rr_2-1-eps-converted-to.pdf}
%         \includegraphics[width=\linewidth]{./rr_2-1_jan25-eps-converted-to.pdf}
%       \end{center}
%     \end{minipage}
    
%     \begin{minipage}{0.5\hsize}
%       \begin{center}
%         %\includegraphics[width=\linewidth]{./rr_3-2-eps-converted-to.pdf}
%         \includegraphics[width=\linewidth]{./rr_3-2_jan25-eps-converted-to.pdf}
%       \end{center}
%     \end{minipage}
%     \\
%     \begin{minipage}{0.5\hsize}
%       \begin{center}
%         %\includegraphics[width=\linewidth]{./rr_4-3-eps-converted-to.pdf}
%         \includegraphics[width=\linewidth]{./rr_4-3_jan25-eps-converted-to.pdf}
%       \end{center}
%     \end{minipage}
           
% \end{tabular}
% \caption{The radiative recombination (RR) rates (to ground state) as a function of temperatures for different elements.
%   The different panels represent different recombination stages (upper left: II to I, upper right: III to II, bottom: IV to III).
%   The same for Fe calculated using analytical method by \ct{Axelrod80} is plotted for comparison.}
%   \label{fig:rr}
% \end{figure*}
%

%========================================================
% figure dr 
%========================================================
\begin{figure*}[t]
  \begin{tabular}{c}
 
   \begin{minipage}{0.5\hsize}
      \begin{center}
       
        \includegraphics[width=\linewidth]{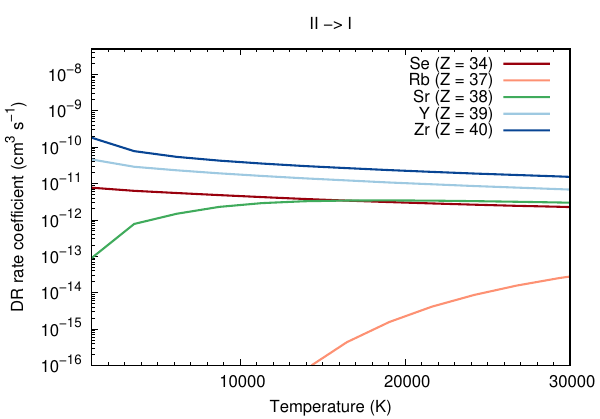}
      \end{center}
    \end{minipage}
    
    \begin{minipage}{0.5\hsize}
      \begin{center}
        
        \includegraphics[width=\linewidth]{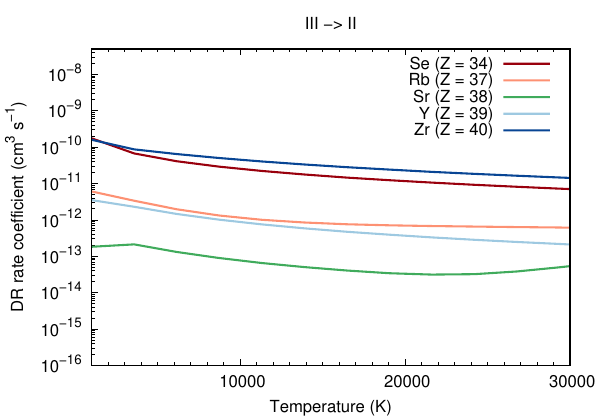}
      \end{center}
    \end{minipage}
    \\
    \begin{minipage}{0.5\hsize}
      \begin{center}
        
        \includegraphics[width=\linewidth]{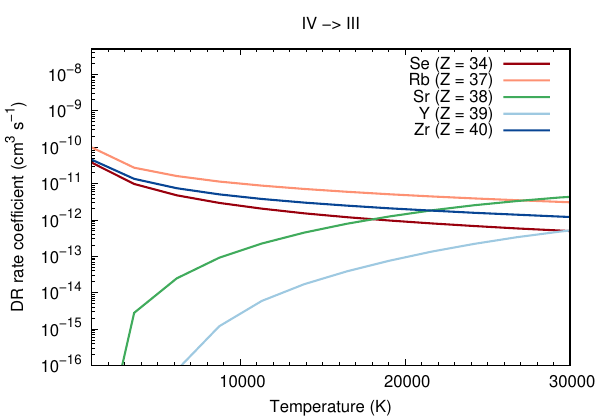}
      \end{center}
    \end{minipage}
          
\end{tabular}
\caption{The dielectronic recombination (DR) rates as function of temperature for the  different elements.
  The different panels represent different recombination stages (upper left: II to I, upper right: III to II, bottom: IV to III).}
  \label{fig:dr}
\end{figure*}

%========================================================

\begin{figure*}[t]
  %\begin{tabular}{c}

    %\begin{minipage}{0.5\hsize}
      \begin{center}

        \includegraphics[width=0.5\linewidth]{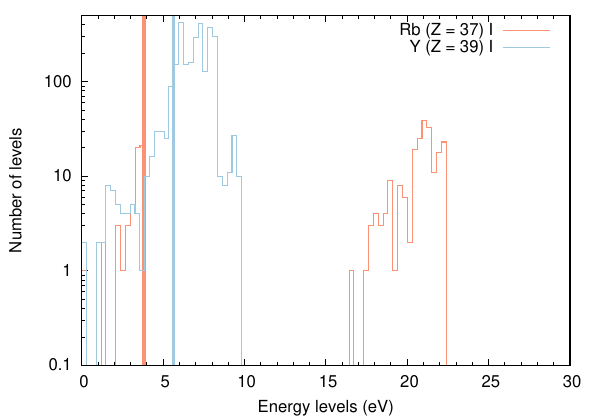}
      \end{center}
    %\end{minipage}
    
    % \begin{minipage}{0.5\hsize}
    %   \begin{center}
    %     \includegraphics[width=\linewidth]{./Rby_auelev_2-1_ai-eps-converted-to.pdf}
    %   \end{center}
    % \end{minipage}  
   
%\end{tabular}
  \caption{Distribution of energy levels of Rb I ({red}) and Y I ({blue}). {The vertical lines are the ionization thresholds for the respective elements.} The autoionizing levels are to the right of the thresholds. {The figure demonstrates that DR can be efficient at low temperatures for Y II to Y I recombination, but not for Rb II to Rb I.}}
  \label{fig:aulev}
\end{figure*}

% %========================================================
\begin{figure*}[t]
  \begin{tabular}{c}

    \begin{minipage}{0.5\hsize}
      \begin{center}
        
        \includegraphics[width=\linewidth]{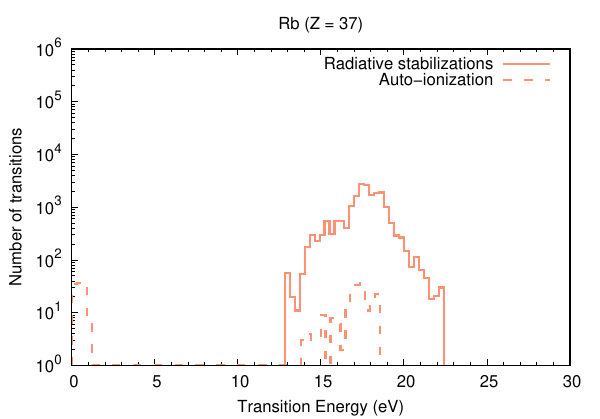}
      \end{center}
    \end{minipage}
    
    \begin{minipage}{0.5\hsize}
      \begin{center}
        \includegraphics[width=\linewidth]{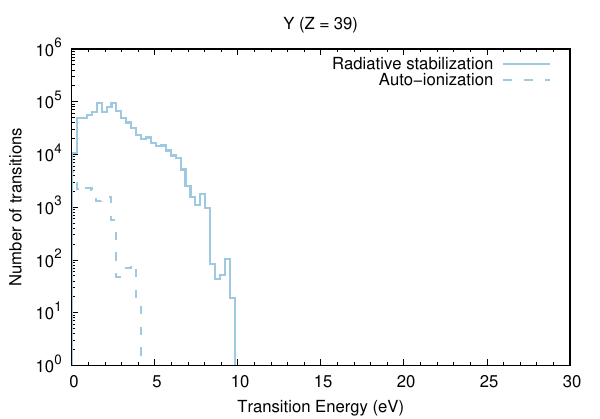}
      \end{center}
    \end{minipage}  
   
\end{tabular}
  \caption{{Histograms of number of transitions from the autoionization levels for neutral Rb I (left) and neutral Y I (right).
  The solid lines represent radiative stabilisation and the dashed lines represent autoionization.}}
  \label{fig:autran}
\end{figure*}
%========================================================
%========================================================
\begin{figure*}[t]
  %\begin{tabular}{c}

    %\begin{minipage}{0.5\hsize}
      \begin{center}

        \includegraphics[width=0.5\linewidth]{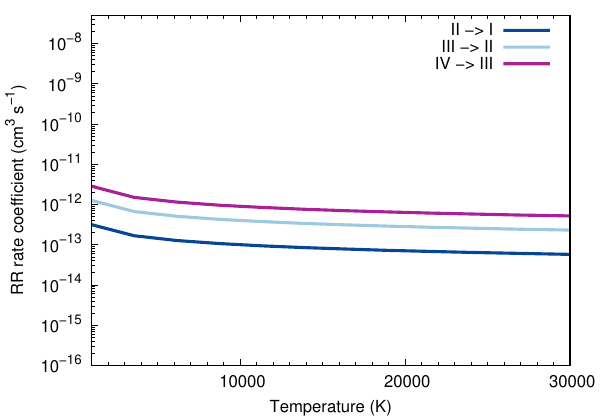}
      \end{center}

  \caption{{The radiative recombination (RR) rates (to ground state) as a function of temperatures for Fe calculated using the analytical method by \ct{Axelrod80} for different ionizations.}}
  \label{fig:rr}
\end{figure*}

%========================================================

%========================================================
% figure dr 
%========================================================
\begin{figure*}[t]
  \begin{tabular}{c}
 
   \begin{minipage}{0.5\hsize}
      \begin{center}
        \includegraphics[width=\linewidth]{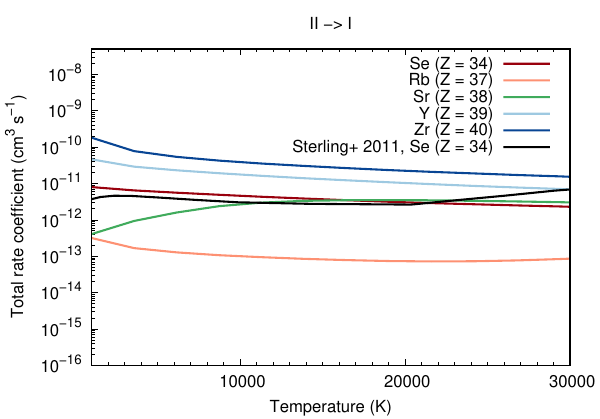}
      \end{center}
    \end{minipage}
    
    \begin{minipage}{0.5\hsize}
      \begin{center}
        \includegraphics[width=\linewidth]{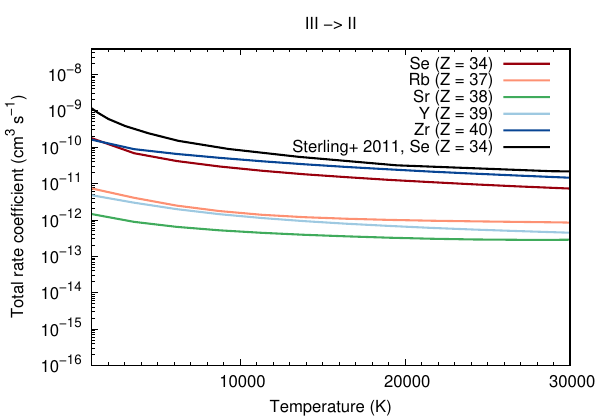}
      \end{center}
    \end{minipage}
    \\
    \begin{minipage}{0.5\hsize}
      \begin{center}
        \includegraphics[width=\linewidth]{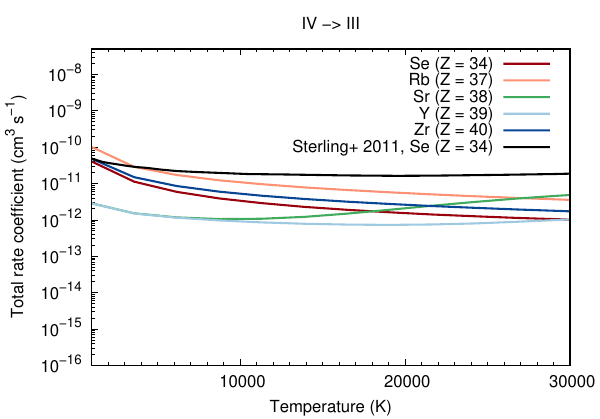}
      \end{center}
    \end{minipage}
    
\end{tabular}
\caption{The total recombination rates (DR total plus RR ground state) as a function of temperature.
{The RR is the value for Fe taken from \ct{Axelrod80}. The black curves are the total recombination rates taken from \ct{Sterling11}, shown for comparison.}
The different panels represent different recombination stages (upper left: II to I, upper right: III to II, bottom: IV to III).}
  %\caption{The Recombination rate for  .}
  \label{fig:tot}
\end{figure*}

%========================================================
\section{Atomic calculation results} \label{sec:rr_dr}
%========================================================
In \ar{fig:eth}, we compare the ionization threshold energies obtained from \texttt{HULLAC} to those provided in the NIST database \ctp{NIST20}.
This comparison serves as an indication of the overall accuracies of the \texttt{HULLAC} results.
The \texttt{HULLAC} results show overall good agreement ($\sim 5$\%) with NIST values across the different ions considered.
The differences are relatively larger for Se I and Sr II ($\sim 15$\%), which is likely attributable to the complex electronic structures of these ions.
For instance, neutral and singly ionized Se (Se I and II) have the nearly half-filled $4p$-shell in their ground configurations ($4s^2\, 4p^4$ and $4s^2\, 4p^3$ for Se I and II, respectively).
Likewise, singly and doubly ionized Sr (Sr II and III) have the completely filled $4p$-shell ($5s\, 4p^6$ and $4p^6$ for Sr II and III, respectively) in the ground configuration. 
The agreement improves for higher ionized species considered, as the electronic configurations becomes relatively simpler.
Take for example, the match between the ionization thresholds approaches near unity for the transitions from the ionization state of III to IV.

%% \subsection{Radiative recombination rates} The RR rate coefficients are shown in \ar{fig:rr}. The values span a wide range of values,
%% from $10^{-15}$ to $10^{-9} \, \rm cm^3\,s^{-1}$, depending on the ion species.
%% The RR rates are generally higher for Se and Rb ions. Previous studies, such as \ct{Hotokezaka21}, use analytical approximations of RR rates for Fe ($Z = 26$) from \ct{Axelrod80},
%% assuming these rates are similar to the rates for the heavier elements as well. 
%% To check the accuracy of this assumption, we overplot these analytical rates (black curves in \ar{fig:rr}) with our detailed calculations. 
%% We find that the detailed rates vary significantly across the temperatures in comparison to these analytical values.
%% This suggests that the use of detailed rates will likely  influence the spectral calculations.

%============================================
\subsection{Dielectronic recombination rates} 
%============================================
\ar{fig:dr} shows the computed DR rate coefficients (in $\rm cm^3 \,s^{-1}$) for various ions as a function of temperature.
We show the DR rate coefficients for the temperature range $T = 1,000 - 30,000$ K, as this is the typical range for nebular-phase kilonovae \ctp{Pognan22a}.
%This is because the temperature in the nebular phase typically lies around $T \sim 10,000$ K (up to $\sim$ first few weeks, \cta{Pognan22a}), and can be increased up to $\sim 30,000$ K (at around $t = 100$ days, depending on the ejecta composition ). 

The dielectronic recombination (DR) rates for different ions considered show different trends with temperature.
We first discuss the DR rates of the ions Rb ($Z = 37$) II and Y ($Z = 39$) II.
For Rb II, the DR rate is negligible at temperatures $T < 15,000$ K, and stays low also at higher ones ($ \sim5\times10^{-14}\, \rm cm^3\,s^{-1}$ at $T \sim 30,000$ K).
On the other hand, for Y II, the DR rate shows a high value of $\sim 5\times10^{-11}\, \rm cm^3\,s^{-1}$ already at $T \sim 1,000$ K. It has an opposite trend with temperature, with a weak decline with higher temperatures.

The DR rates are determined both by the dielectronic capture rates and the fraction (branching ratio) that radiatively stabilizes.
To understand whether dielectronic capture is possible at a certain temperature,
we plot the distribution of energy levels, in \ar{fig:aulev}, and in addition also the threshold energies.
For Rb I, the autoionization levels lie at relatively high energies ($E > 16$ eV), well beyond the threshold. These can be populated only at high electron temperatures. 
On the other hand, Y I has a relatively dense distribution of autoionizing energy levels spanning from threshold to $E \sim 10$ eV, possible to populate also at low electron temperatures.
%
%This explains the negligible DR rates of Rb II at lower temperatures. Since the autoionization levels start becoming populated at relatively high temperatures, there is a gradual increase in the DR rate.
%On the contrary, the autoionizing levels of Y I lie at relatively low energies, which makes it easier to populate them even at low temperatures.
%Hence, the value of the DR rate for Y II to Y I recombination is high already at low temperatures. We see a  slightly decreasing trend following the trailing tail of the Boltzmann distribution.

The values of the DR rates are also quite different between the two ions at all temperatures considered.
Going back to our examples, at a temperature of $T =  20,000$ K, the difference in the DR rates between  Rb II and Y II are almost 3 orders of magnitudes, with Y having the higher value.
{This is because the energy level density is relatively high for Y I, due to the presence of the $4d$-shell. Also, the autoionization levels are low-lying. The higher density of the energy levels and the relatively low-lying autoionizing energy levels make the number of transitions possible to radiatively stabilize from autoionization levels relatively higher (\ar{fig:autran}). }
% This is because the energy level density is relatively high for Y I, due to the presence of the $4d$-shell.
% Also, the autoionization levels are low-lying. Hence, the number of transitions possible to radiatively stabilize from autoionization levels are relatively higher. %(\ar{fig:autran}). 

The variations in the DR coefficients in the other ions can be understood in the similar way. Around the temperature $T = 10,000$ K,
DR rate coefficients for recombining from the ionization state of II to I
vary between $10^{-12} - 5\times10^{-11} \, \rm cm^3\,s^{-1}$ for different elements considered.
For recombination from the ionization state III to II the range spanned is $10^{-13} - 5\times10^{-11} \, \rm cm^3\,s^{-1}$, and for IV to III it is $2\times10^{-15} - 10^{-11} \, \rm cm^3\,s^{-1}$.
One noteworthy result is that out of the different elements across different ionizations, Zr (navy blue curve in \ar{fig:dr})
shows relatively higher values for almost the whole temperature range considered.
This is because Zr has a dense energy level structure (not shown) for both the recombining and the recombined ion due to the presence of the $4d$-shell in its ground state.
Also, the autoionizing levels are relatively low-lying in this case, making it possible to populate at relatively lower temperatures. 

%============================================
\subsection{Total recombination rate}
%============================================
{The total rate coefficients include both RR and DR contributions.
The RR rate is estimated for Fe using the analytical method by \ct{Axelrod80} for diferent ions (see \ar{fig:rr}) and assumed to represent the heavier elements as well. This treatment has also been used in \ct{Hotokezaka21}.}
The total recombination rate range from {$10^{-13}$ to $2\times10^{-10}\, \rm cm^3\,s^{-1}$} for the different ions (\ar{fig:tot}).
In most of the cases, there is significant temperature-dependency, especially at lower temperatures ($T < 10,000$ K),
contrary to assumptions of a temperature-independent value of $10^{-11}\, \rm cm^3\,s^{-1}$ used in earlier spectral calculations \ctp{Pognan23}. 
This assumption might affect the spectral models from previous works, an issue we will examine in more detail in the next section.

In \ar{fig:totT}, we summarize the total rate coefficients for various ions over a temperature range of $T = 1,000$ to $11,000$ K,
which is crucial for understanding kilonova nebulae a few weeks after the event \ctp{Pognan23}.
For transitions from singly ionized to neutral ions (II to I), Se, Y, and Zr has the highest total rates for the recombination
(ranging between {$\sim 2\times10^{-10}\, \rm\  cm^3\,s^{-1}$ to $\sim 10^{-11}\, \rm cm^3\,s^{-1}$),}
while Rb has the smallest ($\sim 10^{-13}\,- 10^{-12} \rm\  cm^3\,s^{-1}$).
For doubly to singly ionized recombination (III to II), {Se and Zr exhibit} the highest recombination rates ($\sim 10^{-10} \rm\  cm^3\,s^{-1}$), 
whereas {Rb, Sr, and Y} show lower values ($\sim 10^{-12}\,- 10^{-11} \rm\  cm^3\,s^{-1}$).
Finally, for triply to doubly ionized recombination (IV to III), the {differences between the}
different ion species considered {decreases, showing values between} $\sim 10^{-11}\,-10^{-10} \, \rm\  cm^3\,s^{-1}$, {with Sr and Y exhibiting relatively lower values.}
%except for Y, which shows a lower value of $\sim 10^{-15}\,-10^{-13} \, \rm\  cm^3\,s^{-1}$.

%% %============================================
%% \subsection{Photoionization cross sections}
%% %============================================
%% We calculate the PI cross-section from the RR cross sections using the Milne relations. We then fit the PI cross sections as a function of incident photon energy using a broken power law.
%% The coefficients for the fitting are provided in \ar {tab:pi}. Note that we consider the cross-section only from ground to ground transitions for the consecutive ions.

%============================================
\subsection{Comparison to previous works}
%============================================
To ensure the reliability of our results, it is important to benchmark the atomic calculation results with other works.
Unfortunately, for the heavier elements the availabilities for the atomic data are limited.
Hence, to get an idea of the uncertainties existing in our calculations, we compare with the {DR rate} coefficients calculated
for Se ionized to I - III by \ct{Sterling11} using \texttt{AUTOSTRUCTURE} \ctp{Badnell86, Badnell97}. 

%For the RR coefficients, the differences between different Se ions vary from $\sim 1-2$ orders of magnitude, the match being worse for lower ions due to their more complicated structures.
%The differences most probably stems from the fact that the RR coefficients are sensitive to the configuration sets used for calculation, which are not tuned between the two calculations.
For the DR coefficients, {the agreement is overall good, considering that the DR cross-sections} are sensitive to the near threshold energy level structures, which are not tuned between the two calculations.
Although there are differences when the {DR} rates are considered individually, comparison of the total recombination rates show a reasonable match for all the ions, 
{with the match between the total coefficients of the ions Se II $\rightarrow$ I, III $\rightarrow$ II, and IV $\rightarrow$ III being within factors of $\sim$ 2, 2.7, and 4, respectively (see \ar{fig:tot}).}
%The deviation is {relatively} larger for neutral Se due to its complicated structure. 
In our radiative transfer calculations, we use the total recombination rates,
hence, the overall degree of uncertainty propagating into the spectral calculations is indicated by these numbers.
A more detailed discussion about the differences between the recombination coefficients between \texttt{HULLAC} and \texttt{AUTOSTRUCTURE} calculations is planned for future work.

%========================================================
\begin{figure*}[t]
  \begin{tabular}{c}
 
   \begin{minipage}{0.5\hsize}
      \begin{center}
        \includegraphics[width=\linewidth]{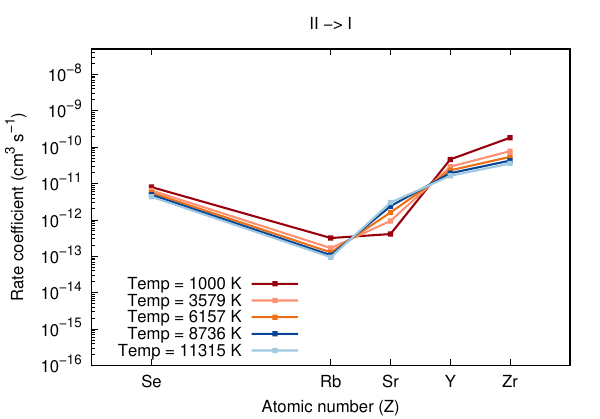}
      \end{center}
    \end{minipage}
    
    \begin{minipage}{0.5\hsize}
      \begin{center}
        
        \includegraphics[width=\linewidth]{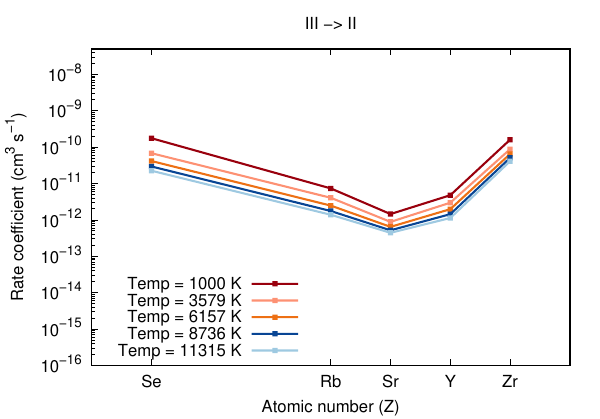}
      \end{center}
    \end{minipage}
    \\
    \begin{minipage}{0.5\hsize}
      \begin{center}
        \includegraphics[width=\linewidth]{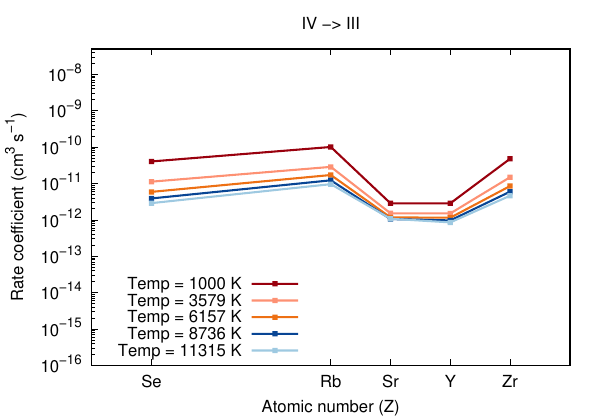}
      \end{center}
    \end{minipage}  
   
\end{tabular}
  \caption{The total recombination rates (DR {calculated from HULLAC} + RR ground {calculated from the analytical formula for Fe from \cta{Axelrod80}}) for different ions at five different temperatures.
  The different panels represent different recombination stages (upper left: II to I, upper right: III to II, bottom: IV to III).}
  \label{fig:totT}
\end{figure*}

\begin{figure*}[h]
%\includgraphics[width=0.49\linewidth]{figures/fig_xe_smarmodel8_10d_smarmodel7_10d.pdf}
%\includegraphics[width=0.49\linewidth]{figures/fig_T_smarmodel8_10d_smarmodel7_10d.pdf}
\includegraphics[width=0.49\linewidth]{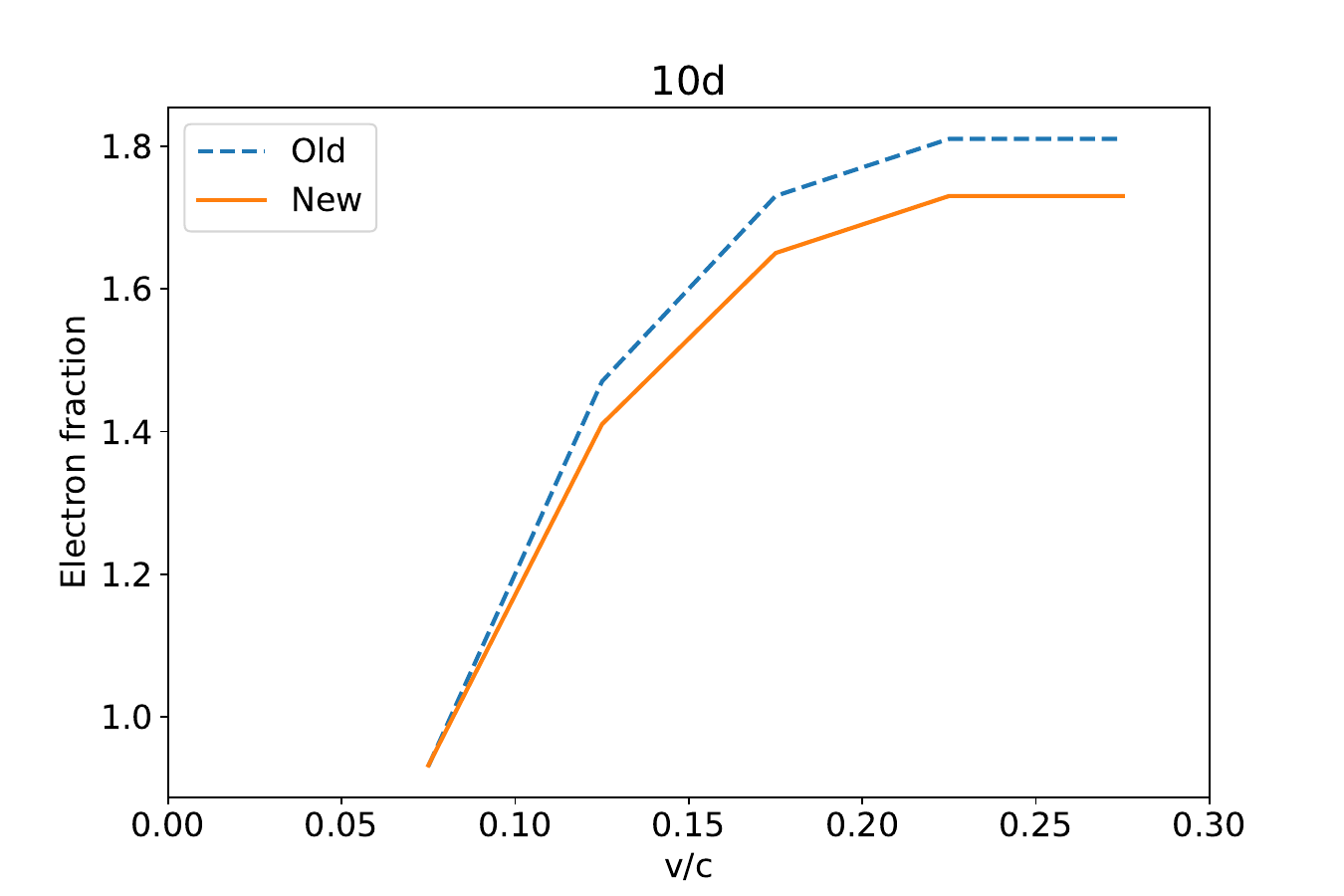}
\includegraphics[width=0.49\linewidth]{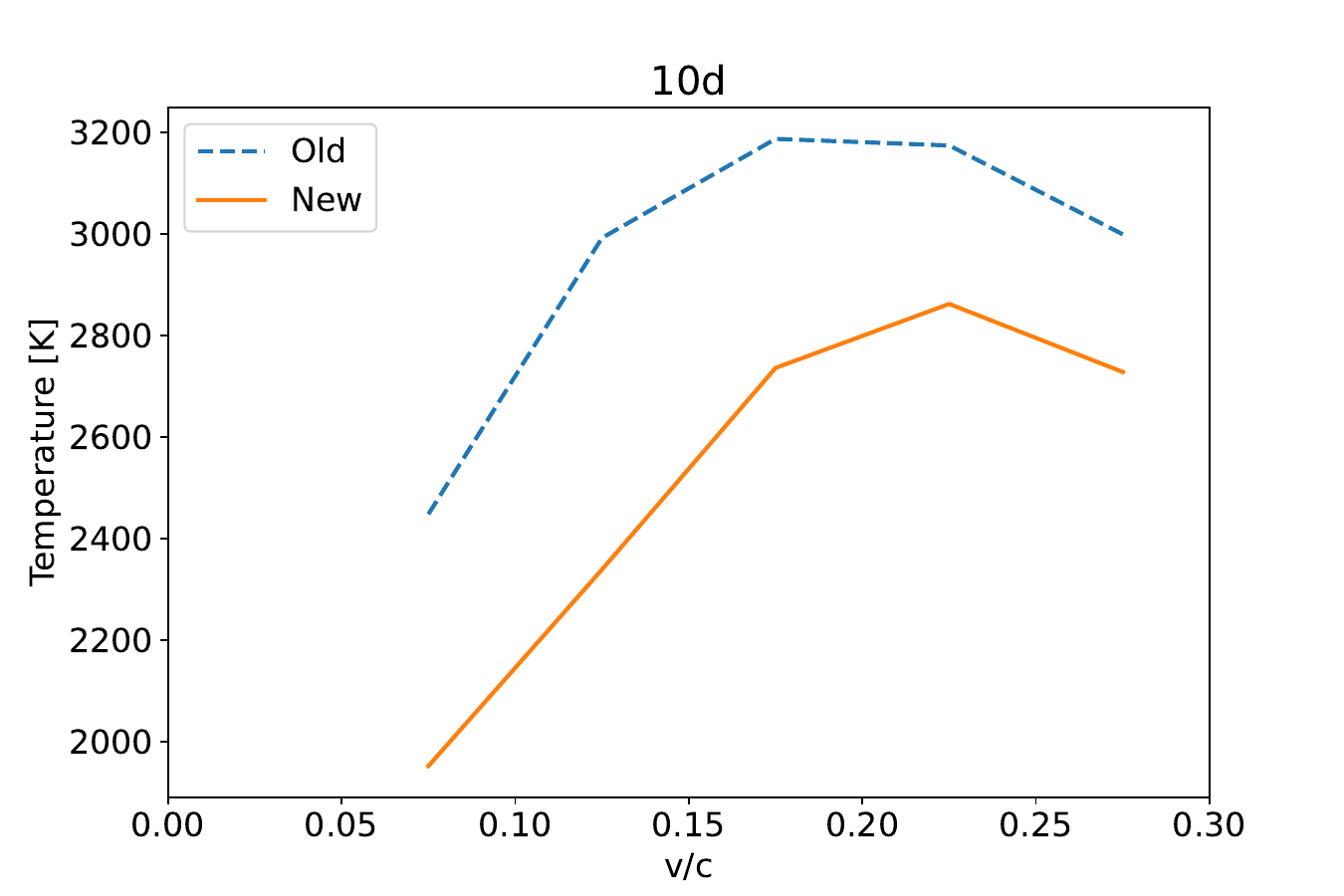}
\caption{Impact on electron fraction (left) and temperature (right) of using the new recombination rates, at $t = 10$ days.}
%and photoionization cross sections
\label{fig:physconds10d}
\end{figure*}
%========================================================
\begin{figure*}[h]
\includegraphics[width=0.49\linewidth]{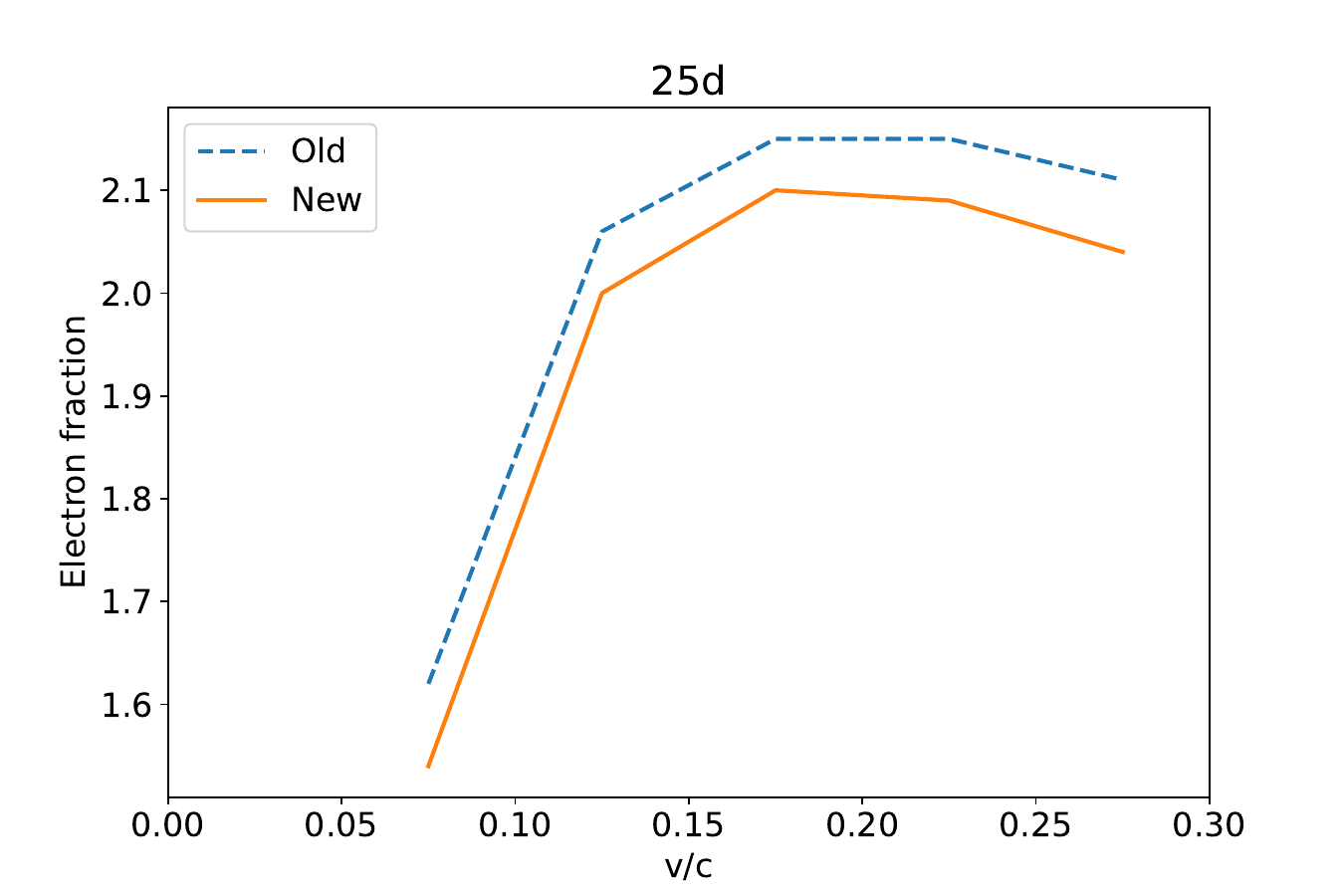}
\includegraphics[width=0.49\linewidth]{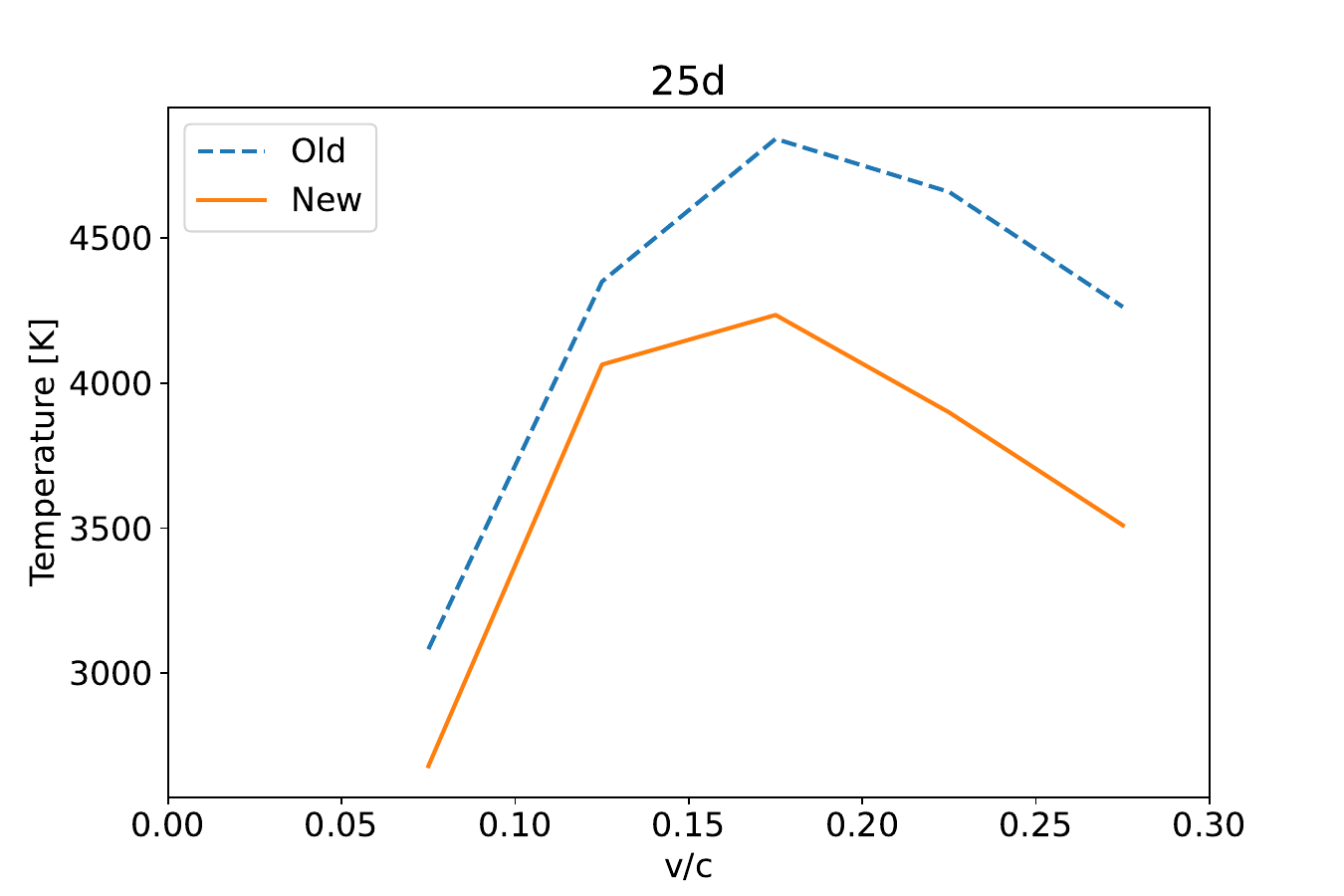}
\caption{Same as \ar{fig:physconds10d}, at 25d.}
\label{fig:physconds25d}
\end{figure*}

%========================================================
\begin{figure*}[h]
\includegraphics[width=0.49\linewidth]{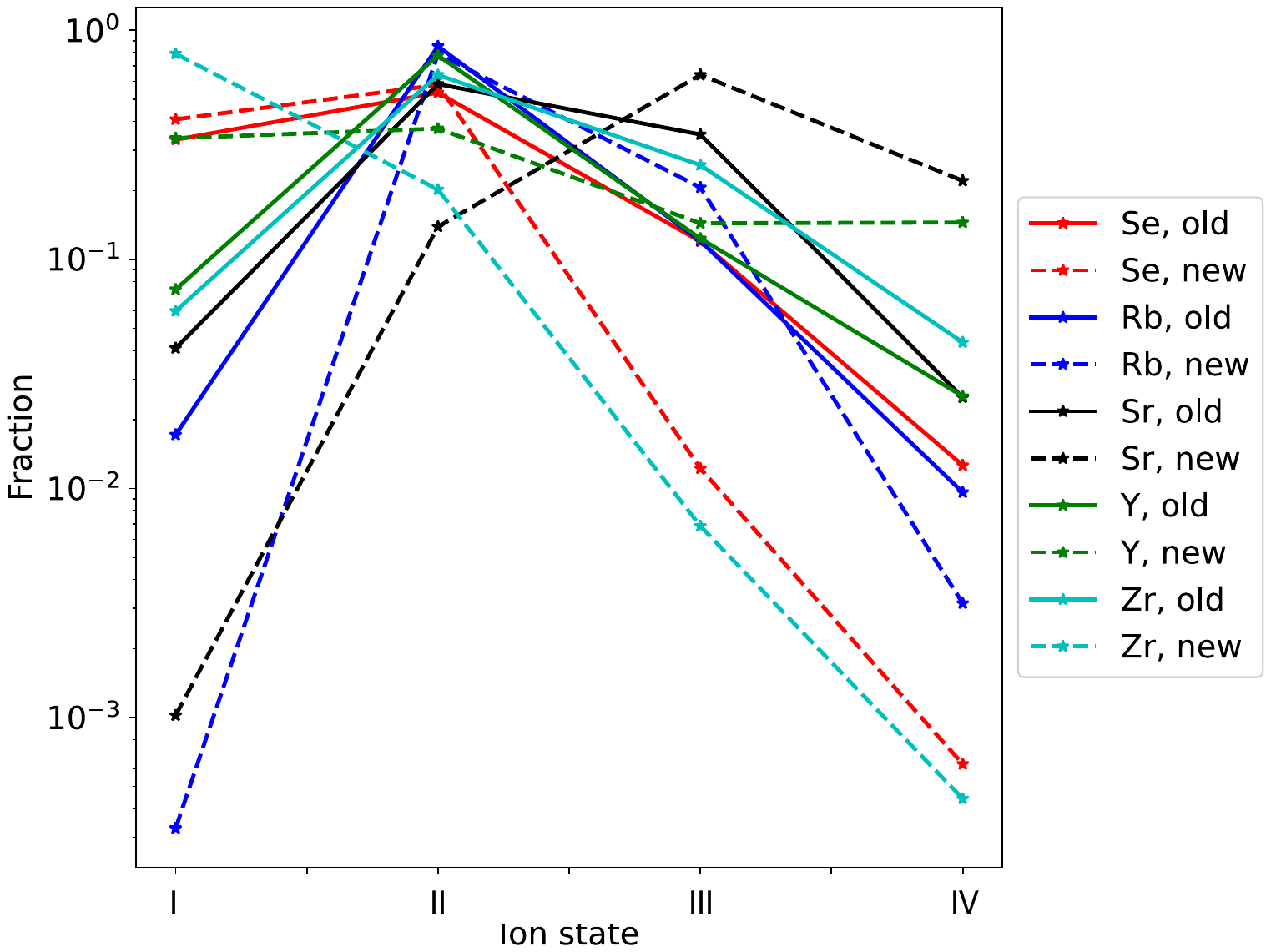}
\includegraphics[width=0.49\linewidth]{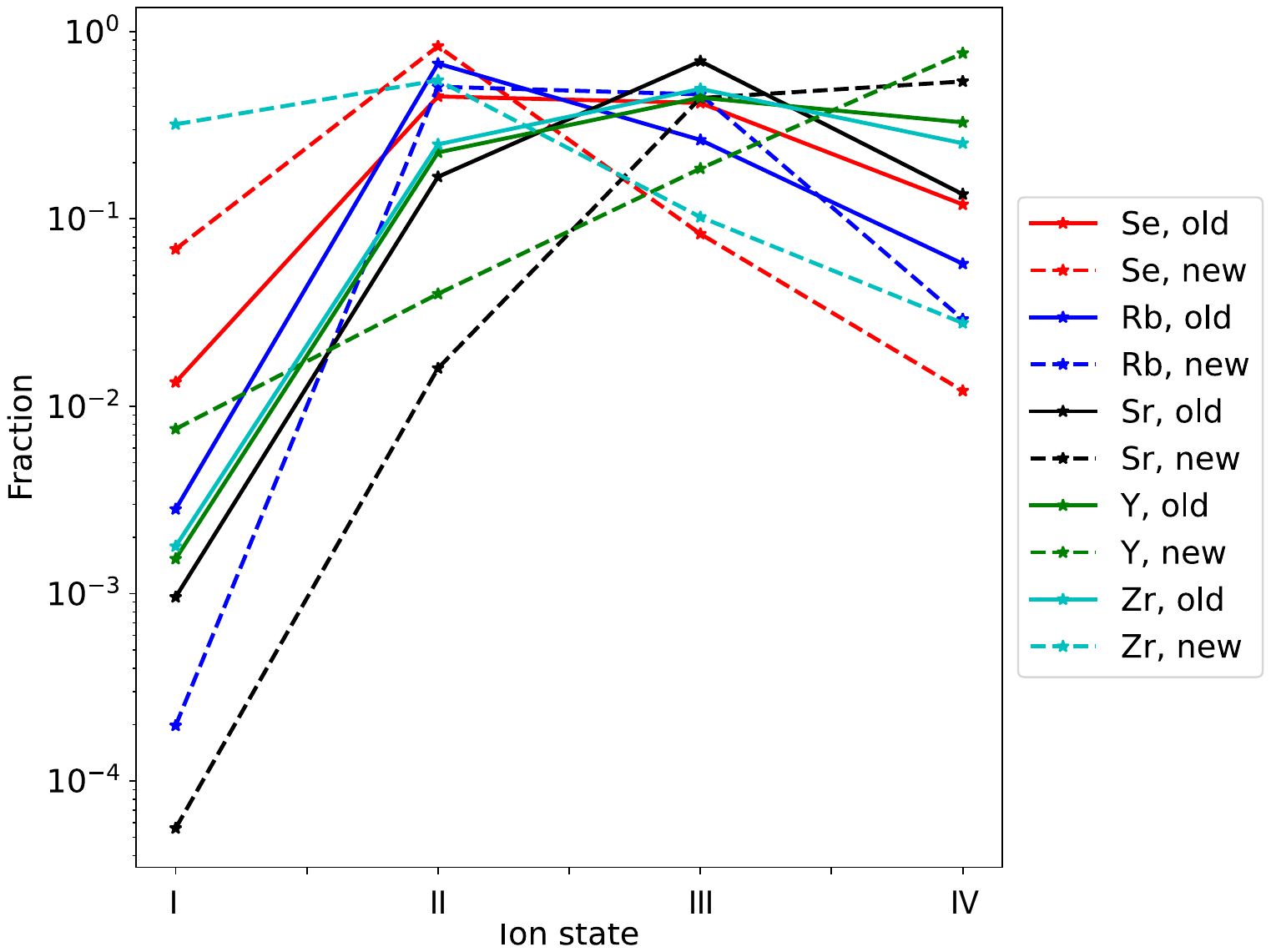}
\caption{Detailed ionization structure changes (innermost zone) for the five elements with new recombination rates, at 10d (left) and 25d (right).}
\label{fig:ionfractions}
\end{figure*}

%========================================================
\begin{figure*}[h]
\includegraphics[width=0.49\linewidth]{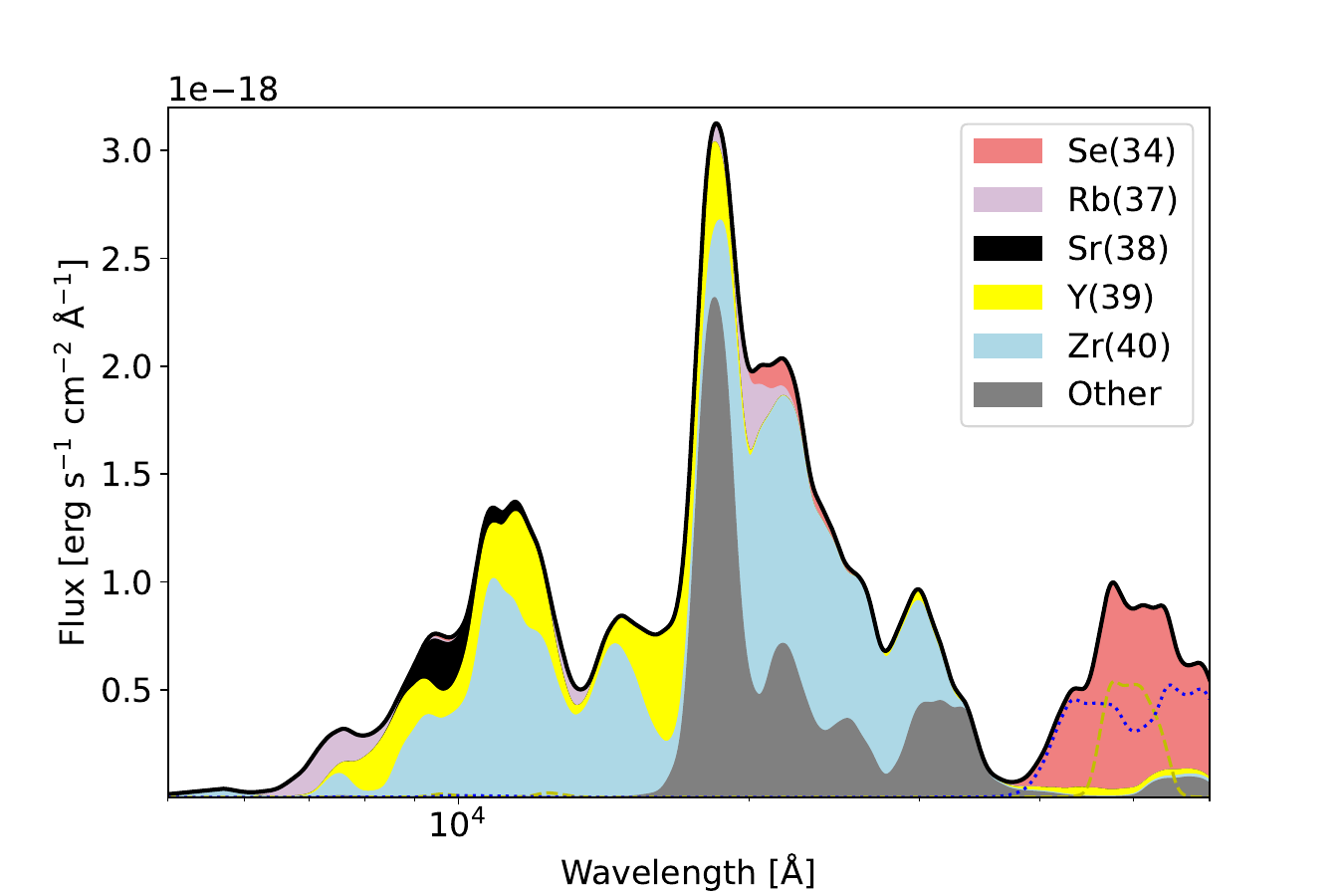}
\includegraphics[width=0.49\linewidth]{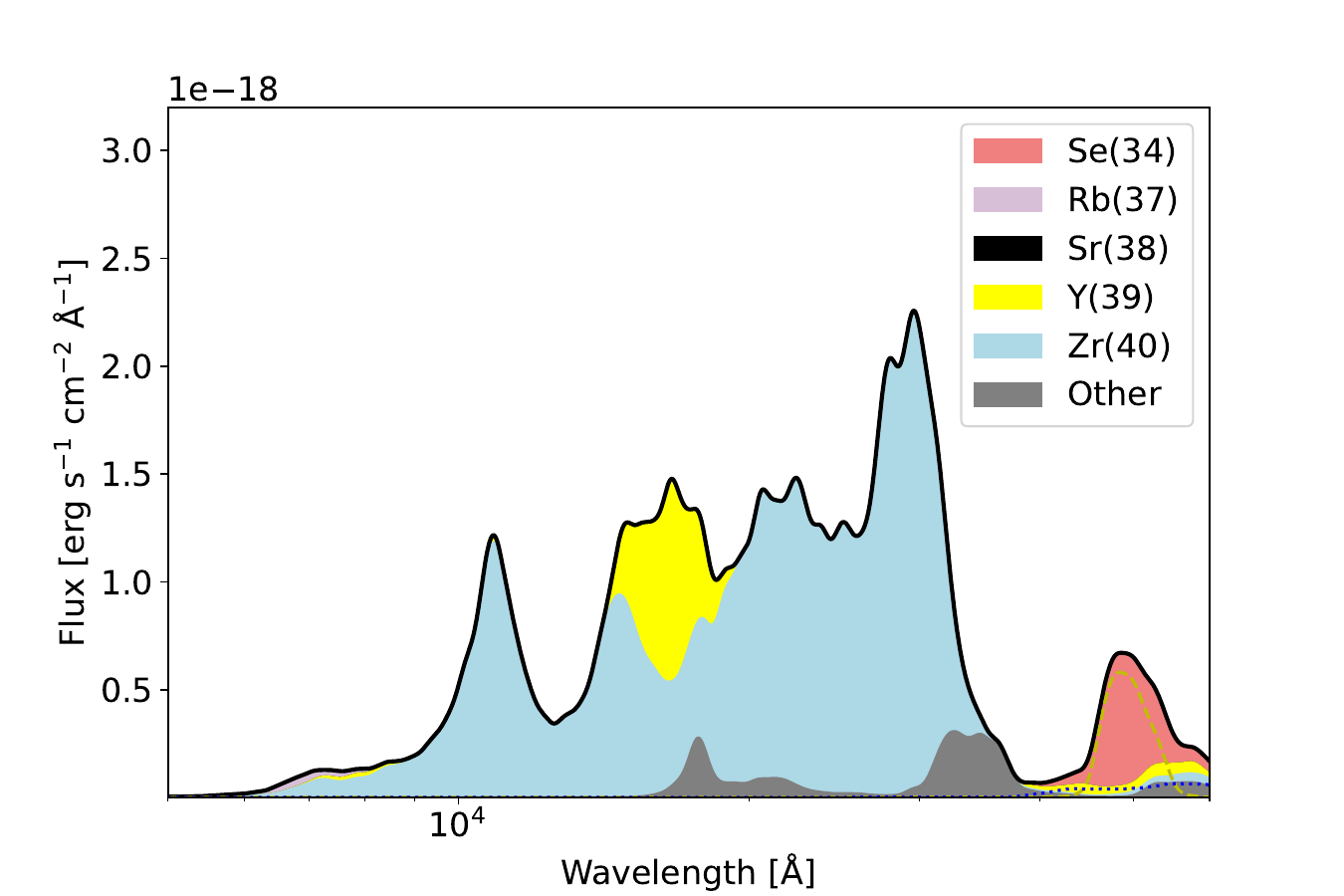}
\caption{Spectrum at 10d of model using old rates (left) and new rates (right), showing the significant impact of the new recombination rates. {Specific contributions by Se I (yellow, dashed line) and Se III (blue, dotted line) are plotted}.}
\label{fig:spectra10d}
\end{figure*}

%========================================================
\begin{figure*}[h]
\includegraphics[width=0.49\linewidth]{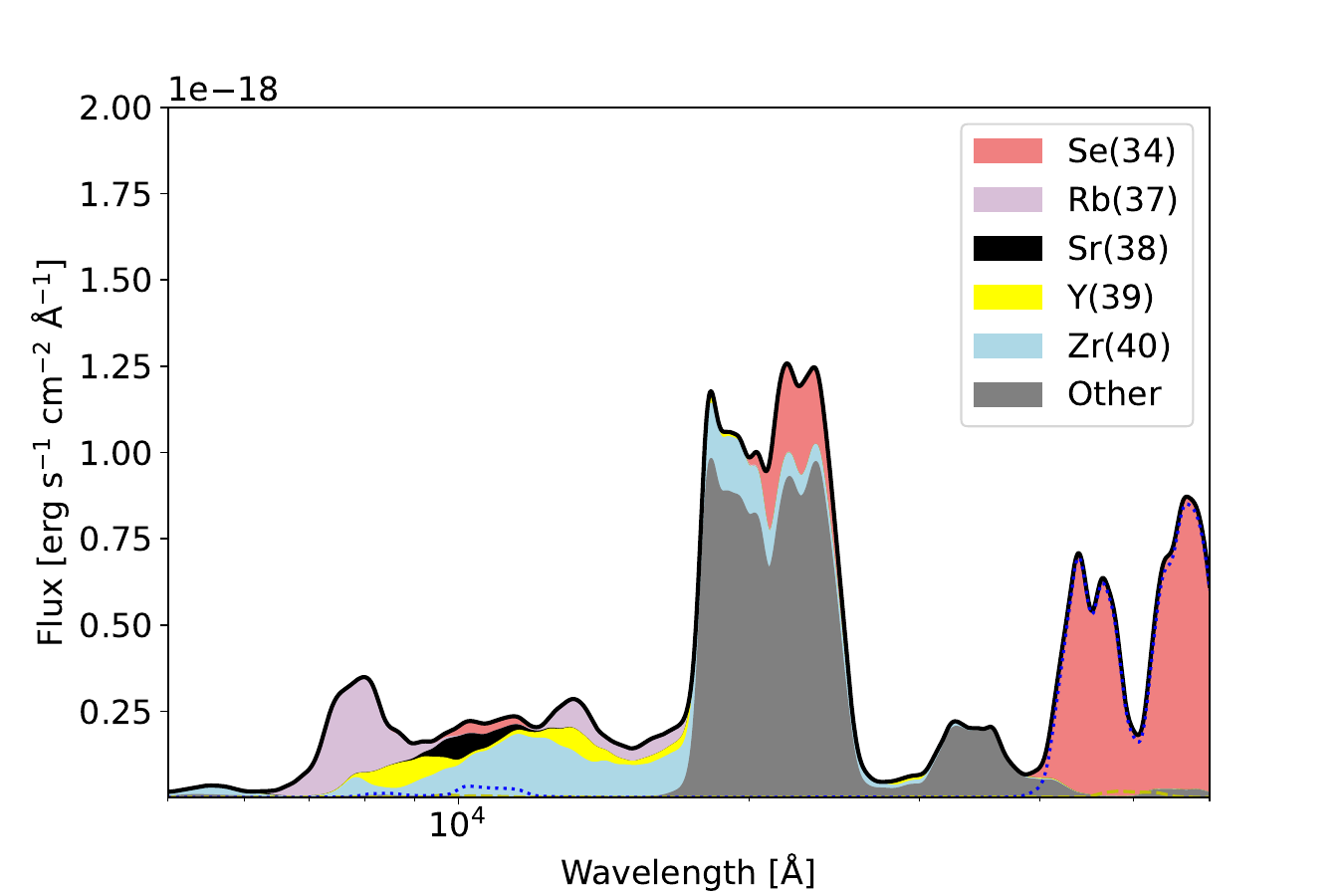}
\includegraphics[width=0.49\linewidth]{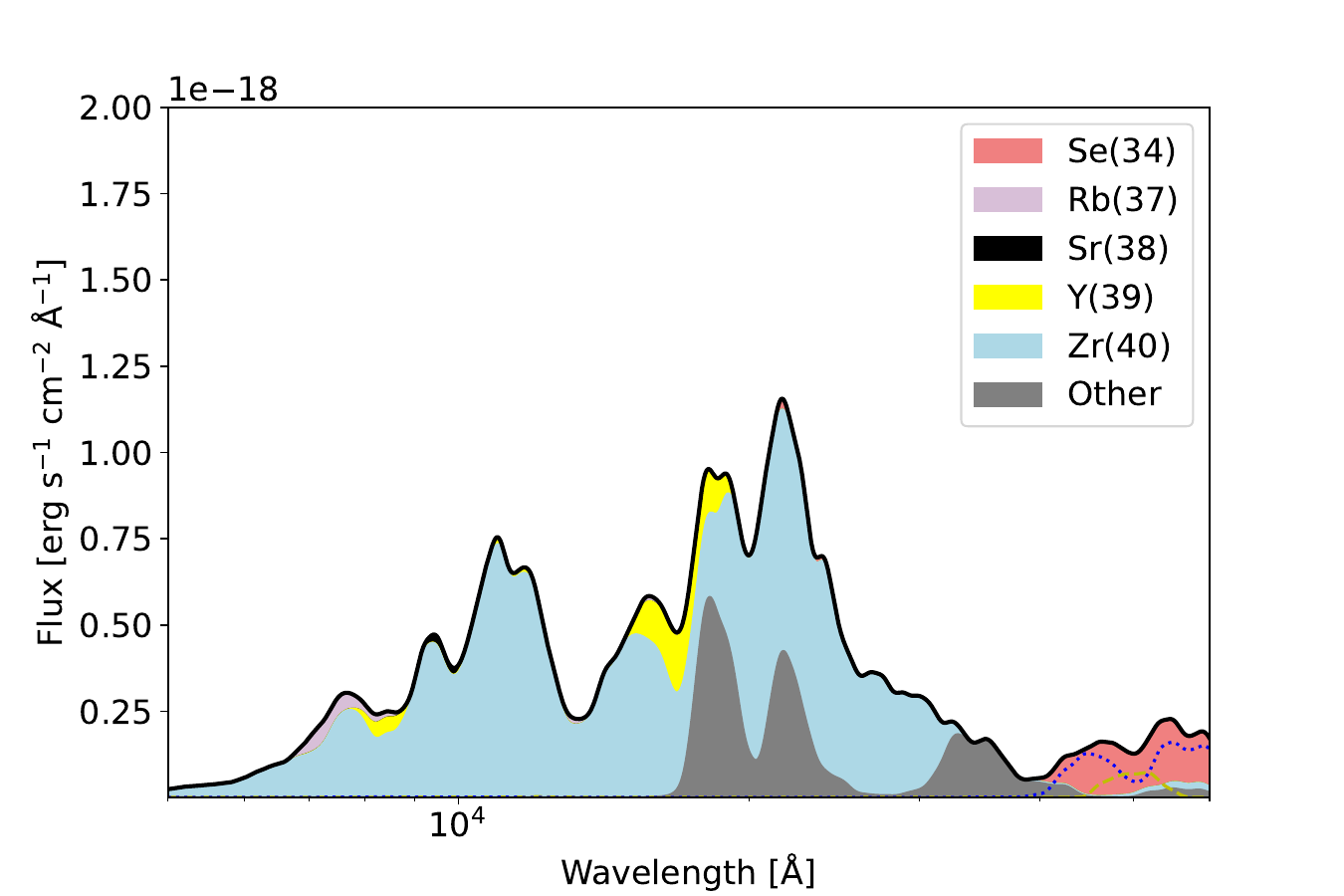}
\caption{Same as \ar{fig:spectra10d}, at 25d.}
\label{fig:spectra25d}
\end{figure*}
%========================================================
\begin{table*}[h]
\centering
\caption{{Percentage contributions by the total ionisation of different elements to the total electron population, in the second innermost zone at 10d and 25d.}}
\label{table:edonors}
\begin{tabular}{ccccc}
\hline
% Total contriutions:
% smarmodel7 (new):  smarmodel8 (old)
Element & Old($t = 10$ days)  & New($t = 10$ days) & Old($t = 25$ days)  & New($t = 25$ days) \\
\hline
Ga ($Z =31$)  & 0.3 & 0.4 & 0.3 & {0.3} \\
Ge ($Z =32$)  & 1.3 & {1.4} & 1.2 & 1.3 \\     
As ($Z =33$)  & 0.1 & 0.1 & 0.1 & 0.1 \\     
Se ($Z =34$)  & 14.9 & {10.2} & 15.1 & {10.3} \\    
Br ($Z =35$)  & 2.5 & {2.5} & 2.7 & {2.7}\\
Kr ($Z =36$)  & 22.8 & {24.1} & 27.3 & {27.8}\\
Rb ($Z =37$)  & 10.7 & {12.2} & 10.2 &  {11.2}\\
Sr ($Z =38$)  & 27.6 & {38.2} & 24.0 & {30.3}\\
Y ($Z=39$)  & {3.8} & {5.4} & {3.8} & {4.4}\\
Zr ($Z =40$) & 15.9 & {5.6} & 18.0 &  {11.5}\\
\hline
\end{tabular}
\end{table*}

%========================================================
\section{Radiative transfer calculation} \label{sec:radtr}
%========================================================

To study the impact of the new atomic data, we compute kilonova spectra in the NLTE phase ($\gtrsim$10d) using the spectral synthesis code \texttt{SUMO} \ctp{Jerkstrand11}.
We compute the spectra at $t = 10$ days and $t = 25$ days, under the steady state approximation.
The details of modelling kilonova at NLTE phase has been discussed in previous work (see \cta{Pognan23}), and hence, we only provide a brief summary of the modelling in \ar{sec:model_spec}.

We use an ejecta with a total mass of 0.05 $M_\odot$ and a power law density profile with an index $n = 4$, similar to that of polar dynamic ejecta \ctp{Kawaguchi20},
distributed within the velocity range of $v_{ej}=0.05c - 0.3c$. We divide the ejecta into five radial zones, spaced linearly with a velocity step of $v_{\rm step} = 0.05c$.

We use a light $r$-process composition, with abundances based on the $Y_e=0.35$ trajectory of \citet{Wanajo14}, 
limiting the composition to the ten elements in the $Z=31-40$ range (which make up $\gtrsim$ 80\% of the mass and provides most contribution to the spectra, \cta{Pognan23}).
We use the radioactive decay and thermalization physics described in \cta{Kasen19} and \cite{Waxman2019}.

For the atomic data related to the ionization structure solutions, {we use the total recombinations rates for $Z=34,\,37-40$ as calculated in this work and described in \ar{sec:rr_dr}. For PI cross-sections, we use hydrogenic values as before.}
%To compensate for the fact that our RR rates consider the transitions between the ground states of the consecutive ions only, we multiply the total RR rate by a factor of 10.
%Furthermore, we assume that the PI cross-section for the whole ground term \ctp{NIST20} is the same as that of the ground level.
%Cross sections for other excited states are treated in the hydrogenic approximation. 
%
We refer to the model using the new recombination rates as the \textit{new} model. We also calculate one model using, as in \cta{Pognan23}, constant recombination rates $10^{-11}$ cm$^3$s$^{-1}$. We refer to this model as the \textit{old} model.

We use a default collision strength of 10 times the \ct{Axelrod80} value {(so $\Upsilon=0.04 g_1 g_2$, where $g_1$ and $g_2$ are the statistical weights of the two levels)} to improve consistency with new r-process calculations \ctp{Bromley23}. For the rest of the atomic data, we mostly follow the treatment in \cta{Pognan23}. However, we update the energy levels of a few low-lying states in selected ions of particular importance, retrieving experimentally verified energy levels from NIST \citep{NIST20}.
For Se I, we update the energies of levels 2-5, for Se II levels 2-5, for Se III levels 2-5, for Se IV level 2, with all Se data from \citet{Moore71}. For Y I, level 2 \citep{Palmer77} is updated. Finally for Y II and Zr III, levels 2-12 \citep{Nilsson91} and levels 2-8 \citep{Reader97} are updated. 

Some commenting is warranted regarding the treatment of bound-free (BF) emission. In detail, RR is associated with a BF emission governed by the specific capture cross-section as function of electron energy. \texttt{SUMO} however does not treat BF emission in that level of detail, as typically only the RR rates (velocity-integrated cross sections) are entered in the atomic data library. The BF emissivity is therefore approximated as a flat emissivity over energies $\chi_i + 1/2kT$ to $\chi_i+3/2kT$, for recombination to level $i$, where $\chi_i$ is the ionization potential \citep{Jerkstrand11}.  

For dielectronic recombination, the bound-free emissivity are {narrow features} ("lines") at energy $\chi_i + a_j$, where $a_j$ is the energy of autoionizing state $j$. However, summing this up over all AI states will give a set of BF lines covering a similar energy range (as only these can be reached by the thermal electrons). As such, the same treatment is done as for RR, with a flat continuum emissivity. It is probably doubtful that the accuracy of the specific energies of the AI states is good enough that a more detailed treatment would be meaningful, {and the continuous Doppler shifting in the comoving frame also contributes to effectively smearing the opacity}. 

%========================================================
\subsection{Spectra at $t = 10$ days}
%========================================================

\ar{fig:physconds10d} compares the ionization and temperature structures using the old and new recombination rates, at $t = 10$ days. \ar{fig:ionfractions}
shows detailed ionization structures for the elements with new recombination rates. The new model is {somewhat less ionized overall (lower $x_e$)}, 
%with a factor $1.3-2$ lower electron fraction,
with the difference between the old and new model increasing with velocity coordinate. In the old model, the most important electron donors (second innermost zone\footnote{At $t = 10$ days, 77\% and 17\% (94\% total) of the total radioactive energy deposition occurs in the innermost two zones - these two zones are therefore the most important to study physical conditions in order to understand the spectral formation.}) are Sr III (19\%), Kr II (11\%), Kr III (9\%), and Zr III (8\%), whereas the new model has leading contributions by {Sr IV (22\%), Sr III (16\%), Kr II (11\%), Kr III (10\%), and Se II (9\%)}. The contributions by the  elements are summarised in \ar{table:edonors}. The overall {somewhat} lower ionization in the new model is a consequence of the new recombination rates {somewhat} more often being higher, than lower, than the canonical $10^{-11}$ cm$^3$ s$^{-1}$ used in the old model (\ar{fig:tot}). {It shows that although $10^{-11}$ cm$^3$s$^{-1}$ is close to an optimal choice, it cannot reproduce the detailed spectral pattern.}
%{However, the results also confirm that if a fixed value needs be chosen (e.g. for other non-calculated elements), $10^{-11}$ cm$^3$s$^{-1}$ is close to an optimal choice.}

The temperatures decrease significantly in the new model, by several hundred K in each zone. 
%$\sim$600 K in the innermost zone and by over 1200 K in the outermost. 
For constant ionic cooling capacity, a lower $x_e$ (as in the new model) would lead to higher temperatures. However, as different $r$-process ions can have very different cooling capacity \citep{Hotokezaka21, Pognan22}, this can be offset by the changes in the ionization structure, which is what occurs here. Taking the innermost zone as an example, the new model has strong cooling by Zr I (61\%), 
%and Se I (20\%)
which is too rare in the old model to be dominant.
There, instead higher ions such as Kr II do much of the cooling - and their lower cooling efficiencies lead to higher temperatures. In this way, recombination rates indirectly affect the temperatures in kilonovae.

\ar{fig:spectra10d} shows the spectral changes induced by the new atomic data at $t = 10$ days. The new model is more heavily dominated by Zr, with strong Zr I cooling and line blanketing extinguishing most of the optical Rb, Y, and Sr flux seen in the old model. {Another noteworthy change is that [Se III] 4.55 $\mu$m have given way to [Se I] 5.03 $\mu$m (left and right panel of \ar{fig:spectra10d}).}
%and [Se III] 5.74 $\mu$m have given way to [Se I] 5.03 $\mu$m.}
%The lower temperatures have moved significant power out of the optical/NIR range  into the MIR (not plotted), showing that even at $t =10$ days, the photometric light curves can be significantly affected by the detailed recombination rates.

%========================================================
\subsection{Spectra at $t = 25$ days}
%========================================================

\ar{fig:physconds25d} compares the ionization and temperature structures using the old and new recombination rates at $t = 25$ days.
The differences are {slightly larger} at this epoch compared to 10d, with the new rates giving $10-20$\% lower electron fractions in the different zones, maintaining a similar profile. Looking at the most important electron donors, the old model has leading contributions (second innermost zone) by Sr III (14\%), Kr III (12\%), Kr IV (10\%), and Zr IV (10\%), while the new model has {Sr IV (25\%), Kr III (12\%), Kr IV (10\%), and Rb III (7\%)}.
%Thus, there is in this case a less dramatic change to the ionization structure for the most important electron-donating ions.
\ar{table:edonors} shows that the overall contributions by different elements to the electron pool changes moderately at 25d. The overall $10-20$\% reduction is driven by less ionized structures of Se and Zr, which is not offset by slightly more ionized structures of {Sr and Rb}.

%Although the temperature in the innermost zone changes moderately between the models, in zones 2-5 the new model is hotter by over 1000 degrees (right panel, \ar{fig:physconds25d}).
{The new model is cooler by 100-500 K, depending on zone}. While the lower $x_e$ will drive temperatures up (fewer electron collisions per unit time), the larger effect is likely, as at 10d, the different cooling abilities of different ions. For example, for Zr, the new recombination rates are factors of 5-10 higher than the canonical value. In the innermost zone, the ionization solution changes from $\rm{I-II-III-IV}=0.0018-0.25-0.49-0.25$ to {$0.32-0.55-0.10-0.028$, and Zr I and Zr II are good coolants}.  
%Looking again at the second innermost zone as an example, in the old model cooling is dominated by Se III, at 23\%. In the new model, the abundance of the doubly ionized state of Se has changed from 48\% to 30\%. The Se III replacement - mainly Se II, is not as good coolant as Se III and the ionization change in selenium therefore contributes to a net heating. 
Similar changes for the other coolants (each can give a positive or negative net change) together adds up for the total change in temperature.

The direct cooling by recombination is unimportant, at $\lesssim$0.1\% of the total cooling in all zones both with old and new rates, at both $t = 10$ days and 25d.
Similarly is photoelectric heating never competitive with the non-thermal heating by large margins.

\ar{fig:spectra25d} shows the spectral changes induced by the new atomic data at $t = 25$ days. The new model is, as at 10d, more heavily dominated by Zr. Zr I is an extremely effective line blocking agent even at low abundances and hence, with strong Zr I line blanketing extinguishing all of most of the Rb, Y, and Sr flux seen in the old model.

%========================================================
\subsection{Discussion} \label{sec:disc}
%========================================================

In this section, we compare our spectral results with those of the previous works. We first discuss the P-Cygni-like feature at $\lambda = 7500 - 7900$ \AA\ seen in the first-week spectra of AT2017gfo. \ct{Sneppen23} propose that Y II causes this feature. Alternatively, \ct{Pognan23} propose that this line is possibly caused by the Rb I through its [2,3]-1 ground state resonance doublet at 7802, 7949 Å. This feature is analogous to the Na I-D doublet in supernovae, often optically thick despite sodium being mostly ionized.

{In our new model at $t = 10$ days, the ionization for Rb I changes significantly. For example, the ionization of Rb in the innermost zone for ions I-II-III-IV changes from $0.017-0.85-0.12-0.0096$ in the old model to $3.3 \times 10^{-4}-0.79-0.21-3.2 \times 10^{-3}$
in the new model (\ar{fig:ionfractions}). 
The significant reduction of Rb I abundance is driven by the very low DR rate coefficient calculated here (negligible at the relevant temperatures). While the RR rate is unknown, the Axelrod recombination formula gives a rate about 2 orders of magnitude below the canonical total rate of $10^{-11}$ cm$^3$ s$^{-1}$ used for the old model - the Rb I abundance drops proportionally. The Rb I line still leaves an imprint - but significantly weaker than before.
%caused by (1) the higher rates for recombination calculated by the IV$\rightarrow$III and III$\rightarrow$II, ($\sim 10^{-10}$ cm$^3$s$^{-1}$, \ar {sec:rr_dr})
%compared the old canonical value ($10^{-11}$ cm$^3$s$^{-1}$), (2) lower rate for II$\rightarrow$I recombination (about two orders-of-magnitudes),
%and (3) high ground state photoionization cross section for Rb I (Rb I having relatively sparse metastable structuring, hence, ground state photoionization is important).
%Combining all the effects mentioned, the neutral Rb abundance plummets to $10^{-6}$ in the new model, causing Rb I lines to disappear (\ar{fig:spectra10d}).
Similarly, contributions by Y II in the optical regime also does not clearly show up in the new model with detailed recombination rates  (\ar{fig:spectra10d}). This occurs by a combination of lower Y II abundance, and changes in physical conditions. This lowers the confidence in identification of either of these elements in AT2017gfo - although it is important to note that that the model we use for this analysis is at $t = 10$ days, whereas the signature is most clearly seen at $t = 3-6$ days in AT2017gfo \ctp{Sneppen23}.} 

Another noteworthy change is in the Se line proposed by \ct{Hotokezaka22} while analyzing the data for nebular spectra of kilonova AT2017gfo by \textit{Spitzer} Space Telescope (at $t = 43$ days).
\ct{Hotokezaka22} concluded that [Se III] should have a strong emission feature around $\lambda = 4.55\ \mu$m, in case the ejecta are dominated by the first peak elements.
{Note that the epoch of calculations does not match the exact epoch of the previous work. However, we take the closest epoch ($t = 25$ days) for discussion.}
In the old model at $t = 25$ days, [Se III] indeed shows this feature {(left panel, \ar{fig:spectra25d})}, which weakens in the new model {(right panel, \ar{fig:spectra25d})}.
In addition to the [Se III] feature, also the [Se IV] line at $\lambda = 2.29\ \mu$m disappears in the new model {(right panel, \ar{fig:spectra25d})}.
Instead the spectra now has a strong [Se I] line at $\lambda = 5.03\ \mu$m {(right panel, \ar{fig:spectra25d})}.
This is due to the fact that the old model has ionization structure (innermost zone at 25d) of Se as $\rm{I-II-III-IV}=0.013-0.45-0.42-0.20$,
whereas the new one has a less ionized one at {$0.07-0.84-0.08-0.012$}, as the new Se recombination rates are a factor 5-10 higher for all ions compared to the canonical value (\ar{fig:tot}). 

%========================================================
\section{Summary and conclusions} \label{sec:conclusion}
%========================================================
{To investigate the nebular phase spectra of kilonovae, we perform detailed calculations of dielectronic (DR) recombination rates 
for} the light $r$-process elements Se ($Z = 34$), Rb ($Z = 37$), Sr ($Z = 38$), Y ($Z = 39$), and Zr ($Z = 40$).
These elements are chosen because they produce potentially strong spectral signatures in light r-process kilonovae \ctp{Hotokezaka22,Gillanders22, Pognan23}.
We consider recombination to ionization states of I - III, as these are most relevant during the nebular phase (e.g., \cta{Pognan23}).

Our results show that around temperature $T = 10,000$ K, DR rate coefficients for recombining from II to I, III to II, and IV to III vary between
$2\times10^{-12} - 5\times10^{-11} \, \rm cm^3\,s^{-1}$, $10^{-13} - 5\times10^{-11} \, \rm cm^3\,s^{-1}$ and $2\times10^{-15} - 10^{-11} \, \rm cm^3\,s^{-1}$, respectively,
for different elements considered. %Furthermore, the RR rate coefficients span a wide range of values, from $10^{-15}$ to $10^{-9} \, \rm cm^3\,s^{-1}$, depending on the ion species. 
%This makes the total rate coefficients, which include both RR and DR contributions, range from $10^{-13}$ to $10^{-9}\, \rm cm^3\,s^{-1}$ for different ions.

Using the new atomic data, we study the spectra of kilonova at nebular phase using the NLTE spectral synthesis code \texttt{SUMO}  \ctp{Jerkstrand11}.
We study a model with total ejecta mass 0.05 $M_\odot$ with a light $r$-process composition ($Z=30-40$). We calculate the spectra at $t = 10$ days and $t = 25$ days.

The new model shows significant changes in the ionization and temperature profiles, and this
in turn changes model spectra at both $t = 10$ and 25 days. We find that the spectra for the new model is more heavily dominated by Zr.
Existence of the strong line blanketing by Zr I extinguishes the signatures of most of the Rb, Y, and Sr flux seen in the old model,
demonstrating that the capacity to correctly interpret $t \gtrsim 10$d KN spectra critically depends on accurate recombination rates.

We compare our models with the models in \ct{Pognan23} and \ct{Hotokezaka22}. 
With the new recombination rates, we do not obtain the P-Cygni feature at $\lambda = 7500 - 7900$ \AA\  \ctp{Sneppen23} proposed by \ct{Pognan23}
to be caused by the Rb I in its [2,3]-1 ground state resonance doublet at 7802, 7949 Å. 
This occurs because the calculated recombination rate of Rb II $\rightarrow$ I is very low, in combination with an increase of neutral Zr in the ejecta which enhances the line blanketing. 
Due to a more neutral ionization structure of selenium with the new recombination rates, we do not obtain the [Se III] emission feature at $\lambda = 4.55\ \mu$m, proposed by \ct{Hotokezaka22},
but instead a strong [Se I] 5.03 $\mu$m line. This shows the importance of the detailed recombination rates for accurate modelling of kilonova spectra at $t \gtrsim$10 days. 
If a future kilonova is detected close enough (e.g., within $\sim 100$ Mpc), observations by the  \texttt{James Webb Space Telescope} will provide opportunities to test these Se line predictions.
%}

\acknowledgements{We acknowledge funding from the European Research Council
(ERC) under the European Union’s Horizon 2020 Research and Innovation Program (ERC Starting Grant 803189 – SUPERSPEC,
PI: Jerkstrand). SB thanks Dr. Daiji Kato from NIFS, Japan for useful discussion regarding \texttt{HULLAC}. JG thanks the Swedish Research Council for the
individual starting grant with contract no. 2020-05467. The computations were enabled by resources provided by the National
Academic Infrastructure for Supercomputing in Sweden (NAISS), and the Swedish National Infrastructure for Computing (SNIC),
at the Parallelldatorcentrum (PDC) Center for High Performance Computing, Royal Institute of Technology (KTH), partially funded
by the Swedish Research Council through grant agreements no. 2022-06725 and no. 2018-05973. This article is dedicated to the memory of Nigel Badnell, who passed
away during the project.}

%========================================================
\bibliography{./references.bib}{}
\bibliographystyle{aasjournal}
\renewcommand{\thetable}{\Alph{section}\arabic{table}}
\renewcommand\thefigure{\thesection\arabic{figure}} 
\setcounter{equation}{0}
\renewcommand\theequation{A\arabic{equation}}
%========================================================
\appendix
%=======================================================
\section{Radiative recombination} 
%=======================================================
The direct process of recombination, RR, occurs when an electron collides with an ion and recombines to a bound state, 
resulting in the emission of a photon.
This process can be described as
\begin{equation}\label{eqn:radrec}
X_{p}^{(i+1)+} + e^- \rightarrow X^{i+}_{b} + h\nu.
\end{equation}
Here $p$ and $b$ represents the parent state of recombining ion and bound state of recombined ions.
The energy of the emitted photon corresponds to the sum of the binding energy of the electron and its initial kinetic energy. 
Since there are no constraints on the kinetic energy of the electron, it is a non-resonant process.

%% \texttt{HULLAC} calculates the radiative recombination rates by obtaining the relativistic dipole transitions
%% between a bound and a continuum level with the condition that the total energy of the continuum level involved is higher than the bound level.
%% The RR cross-sections are provided as a function of the electron energy,
%% which are then convolved with the Maxwell-Boltzmann velocity distribution for a given temperature to get the RR rates in $\rm {cm^3\, s^{-1}}$.
%% Note that our calculations assume the transitions involving only the ground states of the consecutive ions.
%% For light ($Z=1-26$) elements that channel typically makes up $\sim 5-20\%$ of the total recombination rate -
%% as such the total rate needed in the NLTE calculations should be multiplied by a factor of $\sim$10.

\section{Photoionization cross sections}
PI processes are the inverse of the recombination process.
It is possible to have both direct and resonant PI, corresponding to the direct RR and resonant DR process, respectively.
We use hydrogenic cross-sections in our radiative transfer simulations.
%calculate only the direct PI cross-sections, using detailed balance.

%Note that we compute the PI cross-sections for the ground-to-ground transitions only.

%========================================================
\section{Modelling nebular spectra} \label{sec:model_spec}
%========================================================
The main source of energy deposition in the KN ejecta comes from the radioactive decay of the heavy elements. 
The non-thermal $\beta$-decay products, i.e., $\gamma$-rays,
electrons/positrons ($e^{\pm}$), and for heavy compositions also $\alpha$-particles and fission fragments, 
generate an ionization cascade that heat, ionize and excite the ejecta via collisional interactions. 
In the tail phase, the ejecta are transparent to $\gamma$-rays,
and the for light compositions the thermalization mainly comes from the electrons/positrons ($e^{\pm}$).
Furthermore, we also consider the energy deposition from $\alpha$-decay, although the same from the fission is ignored.
The thermalisation efficiencies for $\beta$ and $\alpha$ decays are taken from \ct{Kasen19} and \ct{Waxman2019}.

Here we want to mention that we consider that the microphysical processes
at nebular phase are fast enough to re-emit the entire energy deposited in the ejecta instantaneously,
i.e., the kilonova nebula is at steady state condition \ctp{Jerkstrand11} throughout the calculations.
\ct{Pognan22a} shows that this is true for most of the neutron star merger ejecta at time $t < 100$ days,
except for the extremely low density or the low power condition in the ejecta.
At later times ($t > 100$ days), the atomic processes becomes slow and the assumption of the steady state breaks down.
However, for our purpose and timescale, assumption of steady state is sufficient.

The energy deposited from radioactive decay also contribute to ionization and excitation.
The ionization structure of the elements in the ejecta is determined by using rate equation, under the assumption of NLTE.
The equation for the rate of change of the ion abundances ($x_{j,i}$ for an element $j$ in ionization state $i$) is given as:

\begin{equation}
\frac{dx_(j,i)}{dt} = \Gamma_{j,i-1} x_{j,i-1} - (\Psi_{j,i} + \Gamma_{j,i}) x_{j,i} + \Psi_{j,i+1} x_{j,i+1}.
\end{equation}
Here $\Gamma_{j,i}$ and $\Psi_{j,i}$ are the total ionization and recombination rate per particle, respectively.
At steady state, we can assume the ionization equillibrium condition to satisfy, leading to set the time derivative as zero.
Hence, we are left with the simpler equation by balancing the ionization and recombination terms:
\begin{equation}
\Gamma_{j,i-1} x_{j,i-1} = \Psi_{j,i} x_{j,i}.
\end{equation}

Note that we assume the ionization occurs via non-thermal electron collisions and photoionization,
whereas the recombination occurs by direct radiative or resonant dielectronic processes.
The ionization and recombination rates generally depend on the ion abundances, and hence,
the equations for determining $x_{j,i}$ are non-linear and they are solved by iteration \ctp{Jerkstrand11}.

The temperature ($T$) of the ejecta at the nebular phase is determined by using the first law of thermodynamics \ctp{Jerkstrand11, Pognan22a}:
\begin{equation}
\frac{dT(t)}{dt} = \dfrac{h(t) - c(t)}{3/2 k_{\rm B}n(t)} - \dfrac{2T(t)}{t}. 
\end{equation}
In the equation, the first term represents the net heating per particle, whereas the second term stems from the cooling from adiabatic expansion of the ejecta.
Here $n(t)$ is the total thermal particle number density, $k_{B}$ is the Boltzmann constant,
$x_e(t)$ is the electron fraction in the ejecta, $h(t)$ and $c(t)$ are the heating and cooling rates per unit volume.
Note that although the ionization cooling should be added as a third term \ctp{Jerkstrand11}, it is negligible for kilonova nebula \ctp{Pognan22a},
and thus, we do not include it here.

This equation is further simplified using the steady state approximation.
At the steady state of nebular phase, the ejecta get continuously heated by the radioactive decay,
as well as get cooled off dominantly by line emission following thermal collisional excitation, recombination, free–free emission, and adiabatic expansion.
If the heating and the temperature of the ejecta increase, the line cooling by collisional excitation is also increased.
Hence, at steady state, the temperature gets balanced and the thermal equillibrium is established,
reducing the equation to the simplified form of $h(t) = c(t)$.

The excitation structure within each ion is calculated by using the detailed balance equation.
The fraction of an element $j$ at ionization state $i$ in excitation state $k \neq k^{'}$ (we omit the symbol  $j$ for simplicity) is given as \ctp{Jerkstrand11}:
\begin{equation}
\begin{aligned}
& \dfrac{d x_{i,k}}{d t} = \Sigma_{k^{'}} x_{i-1, k^{'}} \Psi_{i-1, k^{'},i,k} + \Sigma_{k^{'}} x_{i+1} \Psi_{i+1, k^{'},i,k} \\
& + \Sigma_{k^{'} \neq k} x_{i, k^{'}} \xi_{k^{'},k} - x_{i,k} \bigg{(} \Sigma_{k^{'} \neq k} \xi_{k, k^{'}} \\
& + \Sigma_{k^{'} \neq k} \Gamma_{i,k,i+1,k^{'}} + \Sigma_{k^{'} \neq k} \Psi_{i, k, i-1,k^{'}} \bigg{)},
\end{aligned}
\end{equation}
where $\xi$ is the internal transition and the other symbols have their usual meaning.
The internal transitions can happen via spontaneous emission, stimulated emission, photoabsorptions, and both non-thermal and thermal collisions. 
Under the steady state approximation, the de-excitation time-scale is shorter in comparison to the evolutionary time.
Hence, we set the time derivative term to zero, simplifying the equation.

In summary, the thermodynamic properties of the ejecta during the nebular phase,
and consequently the nebular spectra of kilonovae are highly dependent on the accurate modeling of detailed microphysical processes. 
Therefore, precise atomic cross-sections and rates are essential for understanding these microphysical processes and for subsequent modeling of the nebular spectra of kilonova.

%========================================================
\end{document}